\documentclass[aps,prd,superscriptaddress,nofootinbib,amsmath,amsfonts,preprintnumbers,notitlepage,10pt,english]{revtex4-1}
\usepackage{amsmath}
\usepackage{amssymb}
\usepackage{babel}
\usepackage{graphicx}
\usepackage{dcolumn}
\usepackage{bm}
\usepackage[figtopcap]{subfigure}

\makeatletter

\@ifundefined{textcolor}{}
{%
 \definecolor{BLACK}{gray}{0}
 \definecolor{WHITE}{gray}{1}
 \definecolor{RED}{rgb}{1,0,0}
 \definecolor{GREEN}{rgb}{0,1,0}
 \definecolor{BLUE}{rgb}{0,0,1}
 \definecolor{CYAN}{cmyk}{1,0,0,0}
 \definecolor{MAGENTA}{cmyk}{0,1,0,0}
 \definecolor{YELLOW}{cmyk}{0,0,1,0}
 }
\usepackage{enumerate}
\usepackage{amsmath}
\usepackage{amsfonts}
\usepackage{amssymb}
\usepackage[utf8]{inputenc}
\usepackage[T1]{fontenc}
\usepackage{mathtools}
\usepackage{wasysym}

\DeclareMathAlphabet{\mathds}{U}{BOONDOX-ds}{m}{n}
\usepackage[dvipsnames]{xcolor}

\usepackage[colorlinks=true]{hyperref}
\usepackage{amsthm}

\theoremstyle{definition}

\theoremstyle{plain}

\newcommand{\dd}{\mathrm{d}}

\newcommand{\vect}[1]{\boldsymbol{#1}}
\newcommand{\p}{\partial}
\newcommand{\omt}{\omega_t}
\allowdisplaybreaks

\@ifundefined{textcolor}{}{%
 \definecolor{BLACK}{gray}{0}
 \definecolor{WHITE}{gray}{1}
 \definecolor{RED}{rgb}{1,0,0}
 \definecolor{GREEN}{rgb}{0,1,0}
 \definecolor{BLUE}{rgb}{0,0,1}
 \definecolor{CYAN}{cmyk}{1,0,0,0}
 \definecolor{MAGENTA}{cmyk}{0,1,0,0}
 \definecolor{YELLOW}{cmyk}{0,0,1,0}
 }
\begin{document}
\title{Neutral compact spherically symmetric stars in teleparallel gravity}

\author{G.G.L. Nashed}%
\email{nashed@bue.edu.eg}
\affiliation{Centre for Theoretical Physics, The British University in Egypt, P.O. Box 43, El Sherouk City, Cairo 11837, Egypt}
\affiliation{Egyptian Relativity Group (ERG), Cairo University, Giza 12613, Egypt}
\author{Amare Abebe}%
\email{amare.abbebe@gmail.com}
\affiliation{Center for Space Research, North-West University, Mahikeng 2745, South Africa}
\author{Kazuharu Bamba}%
\email{bamba@sss.fukushima-u.ac.jp}
\affiliation{Division of Human Support System, Faculty of Symbiotic Systems Science, Fukushima University, Fukushima 960-1296, Japan}

\date{\today}
\begin{abstract}

We present novel neutral and uncharged solutions that describe the cluster of Einstein in the teleparallel equivalent of general relativity (TEGR). To this end, we use a tetrad field with non-diagonal spherical symmetry which gives the vanishing of the off-diagonal components for the gravitational field equations in the TEGR theory. The clusters are calculated by using an anisotropic energy-momentum tensor. We solve the field equations of TEGR theory, using two assumptions: the first one is by using an equation of state that relates density with tangential pressure while the second postulate is to assume a specific form of one of the two unknown functions that appear in the non-diagonal tetrad field. Among many things presented in this study, we investigate the static stability specification. We also study the Tolman-Oppenheimer-Volkoff equation of these solutions in addition to the conditions of energy. The causality constraints with the adiabatic index in terms of the limit of stability are discussed.
\end{abstract}

\maketitle
\section{Introduction}
To investigate the importance of astrophysics or astronomy with gravitational waves, the theory of General Relativity (GR) plays an essential role in astrophysical systems like compact objects and radiation with high energy usually from strong gravity field around neutron stars and black holes \cite{Shekh:2019msk}.

Recently, observations show that our universe is experiencing cosmic acceleration. The existence of a peculiar energy component called dark energy (DE) controlling the universe is guaranteed by many observations including type Ia supernovae (SNeIa), the Wilkinson Microwave Anisotropy Probe (WMAP) and the Planck in terms of the cosmic microwave background (CMB) radiation, the surveys of the large-scale structure (LSS) \cite{1999ApJ...517..565P,Spergel:2006hy,10.1046/j.1365-2966.2003.07063.x,Eisenstein_2005, Aghanim:2018eyx}.
In terms of an equation of state for dark energy, $p=\omega\rho$, when $\omega<-1/3$ the accelerated expansion is realized, when $-1/3<\omega<-1$ we have quintessence regime, when $\omega<-1$ we have a phantom regime, and when $p=-\rho$ we have a gravastar (gravitational vacuum condensate star) \cite{mazur2001gravitational,Mazur9545,USMANI2011388,Rahaman:2012wc,RAHAMAN2012319,Bhar2014,Yousaf:2019zcb}.
Explanations for the properties of DE have been proposed; among those are:
1) Modifications of the cosmic energy by involving novel components of DE like a scalar field including quintessence \cite{Copeland:2006wr,Papantonopoulos:2007zz}.
2) Modifications of GR action to derive different
kinds of amendment theories of gravity like $f(T)$ gravity
\cite{Bengochea_2009,Cai:2015emx,Bamba:2012cp,Saha_2018}, where $T$ is the torsion scalar in teleparallelism;
$f(R)$ gravity \cite{DeFelice2010,Nojiri:2010wj,Capozziello:2011et,Nojiri:2017ncd,Capozziello:2010zz,Bamba:2015uma,2010deot.book.....R} with $R$ the scalar curvature; $f(G)$ gravity with $G$ the Gauss-Bonnet invariant \cite{Cognola:2006eg}; $f(R,{\cal T})$ gravity, where ${\cal T}$ the trace of the energy-momentum tensor of matter \cite{Harko:2011kv}, etc.
Teleparallel equivalent of general relativity (TEGR) is another formulation of GR whose dynamical variables are the tetrad fields defined as ${l^i}_\mu$. Here, at each point $x^\mu$ on a manifold, $i$ is the orthonormal basis of the tangent space, and $\mu$ denotes the coordinate basis and both of the indices run from $0\cdots 3$. In Einstein's GR the torsion is absent and the gravitational field  is described by curvature while in TEGR theory, the curvature is vanishing identically and the gravitational field is described by torsion \cite{ARCOS_2004,Sotiriou_2011,Camera_2013,2013tgif.book.....A,Sahlu:2019jmy,Horvat_2015,Liddle_1994}. Fortunately, the two theories describe the gravitational field equivalently on the background of the Lagrangian up to a total divergence term \cite{Nashed:2008fm,Nashed:2019zmy}.

The Einstein's cluster \cite{10.2307/1968902} was presented in the literature at the beginning of the last century to discuss stationary gravitating particles, each of which move in a circular track around the center for them in the influence of the effect from the gravity field. When such particles rotate on the common track and have the same phases, they constitute a shell that is named ``Einstein's Shell''. The construction layers for the Einstein's shell form the Einstein's Cluster. The distribution of such a particle has spherical symmetry and it is continuous and random. These particles have the collision-less geodesics. When the gravitational field is balanced by the centrifugal force the above systems are called static and are in equilibrium. A thick matter shell with the spherical symmetry is constituted by the procedure described above. The resultant configuration has no radial pressure and there exists only its stress in the tangential direction. There are many studies of the Einstein clusters in the literature \cite{10.2307/78530,1968ApJ...153L.163Z,10.1093/mnras/114.6.628,Comer_1993}. For the spherically symmetric case the energy-momentum tensor has anisotropic form, i.e. ${T^0}_0=-\rho$,  ${T^r}_r=p_r$, and ${T^\theta}_\theta={T^\phi}_\phi=p_t$, where ${T^\mu}_\nu$ is the matter energy-momentum tensor, $\rho$ is the energy density, $p_r$ and $p_t$ are the radial and tangential pressures, respectively.
By using the junction condition it can be found that the pressure in the radial direction vanishes for the Einstein's clusters. Recently, the compact objects filled with fluids with their anisotropy have been attracted and their structure and evolutional processes have been studied \cite{10.2307/3560113,Chaisi_2005,PhysRevD.69.084026,Abreu_2007,Thirukkanesh_2008,Maurya2015,PhysRevD.91.044040,Kalam2012,PhysRevD.99.044029,Bhar2017,Bhar2016,Newton_Singh_2017,B_hmer_2006,Andr_asson_2009}. It is the aim of this study to apply a non-diagonal tetrad field that possesses spherical symmetry to the non-vacuum equation of motions of TEGR theory and try to derive novel solutions and discuss their physical contents.

The arrangements of this paper are the followings. In Sec. \ref{S2} we explain the basic formulae in terms of TEGR. In Sec. \ref{S3} the gravitational field equations for the TEGR theory in the non-vacuum background are applied to a non-diagonal tetrad and the non-zero components in terms of these differential equations are derived. The number of the differential equations with their non-linearity is found to be less than the number of unknowns. Therefore, we postulate two different assumptions and derive two novel solutions in this section. In Sec. \ref{S4} we discuss the physical contents of these two solutions and show that the second solution possesses many merits that make it physically acceptable. Among these things that make the second solution physically acceptable is that it satisfies the energy conditions, the TOV equation is satisfied, it has static stability and its adiabatic index is satisfied.
In Sec. \ref{S5} discussions and conclusions of the present considerations are given.

\section{Basic Formulae of Teleparallel Equivalent of General Relativity (TEGR)}\label{S2}
In this section we describe the basic formulae of TEGR.
The tetrad field ${l^i}_\mu$, covariant, and its inverse one $l_i{}^\mu$, contravariant, play a role of the fundamental variables for TEGR. These quantities satisfy the following relation
\begin{equation} \label{tet}
    \vect{l_\nu} = l^i{}_\nu \vect{l_i}, \qquad \vect{l_i} = l_i{}^\nu \vect{l_\nu}.
\end{equation}
Based on the tetrads the metric tensor is defined by
\begin{equation}
     g_{_{_{\beta \alpha}}} = \eta_{ij} l^i{}_\mu l^j{}_\nu = \Vec{l}_\mu \cdot \Vec{l}_\nu\, .
    \label{met}
\end{equation}
Here $\eta_{a b}$ denotes the Minkowski spacetime and it is given by $\eta_{a b}=diag(-1,+1,+1,+1)$. Moreover, $\Vec{l}_\mu$ is the co-frame.

Using the above equations one can easily prove the following identities:
\begin{subequations}
\begin{align}
  \eta_{ij} &= g_{\mu\nu} l_i{}^\mu l_j{}^\nu=\Vec{l}_i \cdot \Vec{l}_j,   &  \eta^{ij}  &= g^{\mu\nu} l^i{}_\mu l^j{}_\nu,  \\
    g^{\mu\nu} &= \eta^{ij} l_i{}^\mu l_j^{\nu},   &  l &= \sqrt{|g|}, \\
    l^i{}_\mu l_i{}^\nu &= \delta^\nu_\mu,   &  l^i{}_\mu l_j{}^\mu &= \delta^i_j\,.
\end{align}
\end{subequations}

With the spin connection the curvature quantity and the trosion one can be written as
\begin{eqnarray}
R^{ij}{}_{\mu \nu} &:=& \p_\mu \omega^{ij}{}_\nu -\p_\nu \omega^{ij}{}_\mu
                    +   \omega^i{}_{s\mu} \omega^{sj}{}_\nu
                    -   \omega^i{}_{s\nu} \omega^{sj}{}_\mu\, , \\
T^i{}_{\mu \nu}     &:=& \p_\mu l^i{}_\nu - \p_\nu b^i{}_\mu
                        + \omega^i{}_{k\mu} l^k{}_\nu - \omega^i{}_{k\nu}l^k{}_\mu,
\end{eqnarray}
where $\omega^{ij}{}_\nu$ is the spin connection.
The matrices with the local Lorentz symmetry, $\Lambda^a{}_b$,
generates the spin connection as
\begin{align}
	\omega^a{}_{b\mu} = \omega^a{}_{b\mu}(\Lambda) = \Lambda^a{}_c \partial_\mu (\Lambda^{-1})^c{}_{b},\quad \eta_{ab}\Lambda^a{}_c\Lambda^b{}_d = \eta_{cd}\,.
\end{align}
The tensors $R^\mu{}_{\nu \rho \sigma}$ and $T^i{}_{\mu\nu}$ are defined as follows:
\begin{enumerate}[(i)]
\item $R^\mu{}_{\nu \rho \sigma} = l_i{}^\mu l_{j\nu} R^{ij}{}_{\rho \sigma},$
\item $l^i{}_\sigma T^\sigma{}_{\mu\nu} = T^i{}_{\mu\nu}\;.$
\end{enumerate}
Using the above data one can define the torsion in terms of the derivative of tetrad and spin connection as
\begin{align}
	T^a{}_{\mu\nu} = 2 \left(\partial_{[\mu}l^a{}_{\nu]} + \omega^a{}_{b[\mu} l^b{}_{\nu]}\right)\,,
\end{align}
where square brackets denote that the pair of indices are skew-symmetric and $\partial_\mu=\frac{\partial }{\partial x^\mu}$.
In the TEGR theory, the spin connection is set to be zero ($\omega^a{}_{b\mu}=0$). Therefore, the torsion tensor takes the form \[T^a{}_{\mu\nu} = \partial_{[\mu}l^a{}_{\nu]}.\]

The TEGR theory is constructed by using the Lagrangian
\begin{align} \label{lag}
	L_{\rm TEGR}= \int \dd^4x\ |l| \left( \frac{1}{2\kappa^2}  T + \mathcal{L}_{\rm m}(g,\Psi) \right)\,,
\end{align}
with $\kappa^2=8\pi$. Here $\mathcal{L}_{\rm m}(g,\Psi)$ is the Lagrangian of matter with minimal coupling to gravitation through the metric tensor written with the tetrad fields. In addition, $T$ is the torsion scalar and it is defined as
\begin{align} \label{tors}
	T =  T^a{}_{\mu\nu}S_a{}^{\mu\nu} =\frac{1}{2} \left(l_a{}^\sigma g^{\rho \mu} l_b{}^\nu + 2 l_b{}^\rho g^{\sigma \mu} l_a{}^\nu + \frac{1}{2} \eta_{ab} g^{\mu\rho} g ^{\nu\sigma} \right) T^a{}_{\mu\nu} T^b{}_{\rho\sigma}\,.
\end{align}
The superpotential $S_a{}^{\mu\nu}$ is defined as \[S_a{}^{\mu\nu} = \frac{1}{2}(K^{\mu\nu}{}_a - h_a{}^\mu T_\lambda{}^{\lambda\nu} + h_a{}^\nu T_\lambda{}^{\lambda\mu}),\] with $K^{\mu\nu}{}_a$ the contortion tensor, expressed by \[K^{\mu\nu}{}_a = \frac{1}{2}(T^{\nu\mu}{}_a + T_a{}^{\mu\nu} - T^{\mu\nu}{}_a)\, .\]
Variation of the Lagrangian (\ref{lag}) with respect to a tetrad $l^a{}_\mu$ yields~\cite{Maluf:2002zc,Nashed_2011}
\begin{align}\label{fe}
	\frac{1}{4}T l_a{}^\mu +  T^b{}_{\nu a} S_b{}^{\mu \nu } + \frac{1}{l}\partial_{\nu}(l S_a{}^{\mu \nu })   &= \frac{1}{2}\kappa^2 \Theta_a{}^\mu\,.
\end{align}
The stress-energy tensor, $ \Theta_a{}^\mu$, is the energy-momentum tensor
for fluids whose configuration has anisotropy and it is represented by
\begin{eqnarray}
&&\Theta_a{}^\mu{}=(p_t+\rho)u^\mu u_a+p_t\delta_a{}^\mu+(p_r-p_t)\xi_a \xi^\mu\, ,
\end{eqnarray}
with $u_\mu$ the time-like vector defined as $u^\mu=[1,0,0,0]$ and $\xi_\mu$ the unit radial vector with its space-like property, defined by $\xi^\mu=[0,1,0,0]$ such that $u^\mu u_\mu=-1$ and $\xi^\mu\xi_\mu=1$.
Here $\rho$ means the energy density,
$p_r$ and $p_t$ are the radial and tangential pressures, respectively.

\section{Neutral compact stars  }\label{S3}
In this section we adopt the gravitational field equation (\ref{fe}) to
the tetrad with its spherical symmetry, which represents a dense compact relativistic star.

Based on the spherical coordinates $(t,r,\theta, \phi)$,
the metric with its spherical symmetry is given by
\begin{equation}
ds^2=-e^{\mu(r)} \,dt^2+e^{\nu(r)}dr^2+r^2d\Omega^2\;, \quad \quad d\Omega^2=(d\theta^2+\sin^2\theta d\phi^2)\,,\label{met1}
\end{equation}
where $\mu(r)$ and $\nu(r)$ are the functions of $r$ in the radial direction. This line element in Eq. (\ref{met1}) can be reproduced from the following tetrad field ~\cite{Bahamonde:2019zea}:
\begin{equation}
l^a{}_{\mu}=\left(
\begin{array}{cccc}
e^{\mu(r)/2} & 0 & 0 & 0 \\
0 & e^{\nu(r)/2} \cos (\phi ) \sin (\theta ) & r \cos (\phi ) \cos (\theta )  & -r \sin (\phi ) \sin (\theta )  \\
0 & e^{\nu(r)/2} \sin (\phi ) \sin (\theta )  & r \sin (\phi ) \cos (\theta )  & r \cos (\phi ) \sin (\theta ) \\
0 & e^{\nu(r)/2} \cos (\theta ) & -r \sin (\theta ) & 0 \\
\end{array}
\right)\label{tet}\,.
\end{equation}
We mention that the tetrad (\ref{tet}) is an output  product of a diagonal tetrad and local Lorentz transformation, i.e., one can write it as
\begin{eqnarray}
&& l^a{}_{\mu}=\Lambda^a{}_b l^b{}_{\mu_{diag}}\nonumber\\
&&\Rightarrow \left(
\begin{array}{cccc}
e^{\mu(r)/2} & 0 & 0 & 0 \\
0 & e^{\nu(r)/2} \cos (\phi ) \sin (\theta ) & r \cos (\phi ) \cos (\theta )  & -r \sin (\phi ) \sin (\theta )  \\
0 & e^{\nu(r)/2}\sin (\phi ) \sin (\theta )  & r \sin (\phi ) \cos (\theta )  & r \cos (\phi ) \sin (\theta ) \\
0 & e^{\nu(r)/2} \cos (\theta ) & -r \sin (\theta ) & 0 \\
\end{array}
\right)\nonumber\\
&&\equiv\left(
\begin{array}{cccc}
1 & 0 & 0 & 0 \\
0 & \cos (\phi ) \sin (\theta ) &  \cos (\phi ) \cos (\theta )  & - \sin (\phi )   \\
0 &  \sin (\phi ) \sin (\theta )  &  \sin (\phi ) \cos (\theta )  &  \cos (\phi )  \\
0 &\cos (\theta ) & -\sin(\theta) & 0 \\
\end{array}
\right)\times\left(
\begin{array}{cccc}
e^{\mu(r)/2} & 0 & 0 & 0 \\
0 & e^{\nu(r)/2} &  0  & 0   \\
0 & 0  &  r  & 0  \\
0 &0 & 0 & r \sin(\theta) \\
\end{array}
\right) \label{tet1}\,.\end{eqnarray}
Using Eq. (\ref{tet}) in Eq. (\ref{tors}) the torsion scalar takes the form
\begin{align}\label{torsc}
  T= \frac{2e^{-\nu} \left[e^{\nu}-e^{\nu/2}(2+r\mu')+1+r\mu'\right]}{r^2}\,.
\end{align}
It follows from Eq. (\ref{torsc}) that $T$ vanishes in the limit $\mu=\nu\rightarrow 0$ unlike what has been studied before in the literature \cite{Bahamonde:2019zea}\footnote{This condition is important since when $\mu=\nu\rightarrow 0$ the line element (\ref{met1}) gives the Minkowski spacetime
 whose torsion has a vanishing value}.
Using Eq. (\ref{torsc}) in the field equations (\ref{fe}) we get
\begin{eqnarray}
 8\pi \rho&=& -\frac{1-e^{-\nu}(1-r\nu')}{r^2}\,,\nonumber\\
8\pi p_r&=&\frac{e^{-\nu}(1-r\mu')-1}{r^2}\,,\nonumber\\
8\pi p_t&=&-\frac{e^{-\nu}\left[2r\mu''+(r\mu'+2)(\mu'-\nu')\right]}{4r}\,,\label{fec}
\end{eqnarray}
where $'$ denotes derivatives with respect to $r$. These differential equations are three independent equations in five unknowns: $\mu$, $\nu$ and $\rho$, $p_r$ and $p_t$. Therefore, we need extra conditions to be able to solve the above system. The extra conditions are the zero radial pressure, namely, $p_r=0$ \cite{Boehmer:2007az,Singh:2019ykp}, and assuming the equation of state (EoS) in terms of the energy density and the tangential pressure. We can represent these conditions as
\begin{align}\label{es}
p_r=0, \qquad \qquad p_t=\omega_t \rho\, ,
\end{align}
where $\omega_t$ is the EoS parameter for anisotropic fluids.

Substituting Eq. (\ref{es}) into (\ref{fec}) we obtain
\begin{eqnarray}\label{sol}
&&\rho=\frac{\omega_t}{2\pi r^2(1+4\omega_t)}, \qquad \mu=4\omt\ln(r)+c_1, \qquad \nu=\ln(4\omt+1), \qquad p_t=\omega_t \rho\, , \nonumber\\
&&\rho=0, \qquad \mu=\ln\Big(\frac{c_3r-c_2}{r}\Big), \qquad \nu=\ln\Big(\frac{rc_3}{c_3r-c_2}\Big), \qquad  p_t=0\,.
\end{eqnarray}
We note that the vanishing of
the reason why the tangential pressure as well as the energy density vanish as in the second set of Eq. (\ref{sol}) is due to the composition of the two unknown functions $\mu$ and $\nu$ that give the Schwarzschild solution.

Another solution that can be derived from Eq. (\ref{fec}) is through the assumption of the unknown function $\mu$ to have the form \cite{Singh:2019ykp} \begin{eqnarray} \label{ass} \mu(r)=\ln(b_0+b_1r^2+b_2r^4)\;. \end{eqnarray}  Using Eq. (\ref{ass}) in (\ref{fec}) we get  the remaining unknown functions in the form
\begin{eqnarray}\label{sol1}
&&\rho=\frac{(3b_1+10b_2r^2)(b_0+b_1r^2+b_2r^4)}{4\pi(b_0+3b_1r^2+5b_2r^4)^2}, \qquad \qquad \qquad \nu=\ln\Big(\frac{b_0+3b_1r^2+5b_2r^4}{b_0+b_1r^2+b_2r^4}\Big)\,, \nonumber\\
&&p_t=\frac{r^2(3b_1[b_1+16b_2r^2]+20b_2{}^2r^4)}{8\pi(b_0+3b_1r^2+5b_2r^4)^2}\;.
\end{eqnarray}

The EoS of the first and second solutions given by Eqs. (\ref{sol}) and (\ref{sol1}) takes the form
\begin{align}\label{es1}
\omega_{t_{1}}=1\,, \qquad \qquad \omega_{t_{2}}=\frac{r^2(b_1+2b_2r^2)}{2(b_0+b_1r^2+b_2r^4)}\,.
\end{align}
The first EoS shows that we have a stiff matter while the behavior of the second EoS is shown in Fig. \ref{Fig:2}\subref{fig:EoS} below.

\begin{figure}
\centering
\subfigure[~Density of the first solution]{\label{fig:dnesity}\includegraphics[scale=0.3]{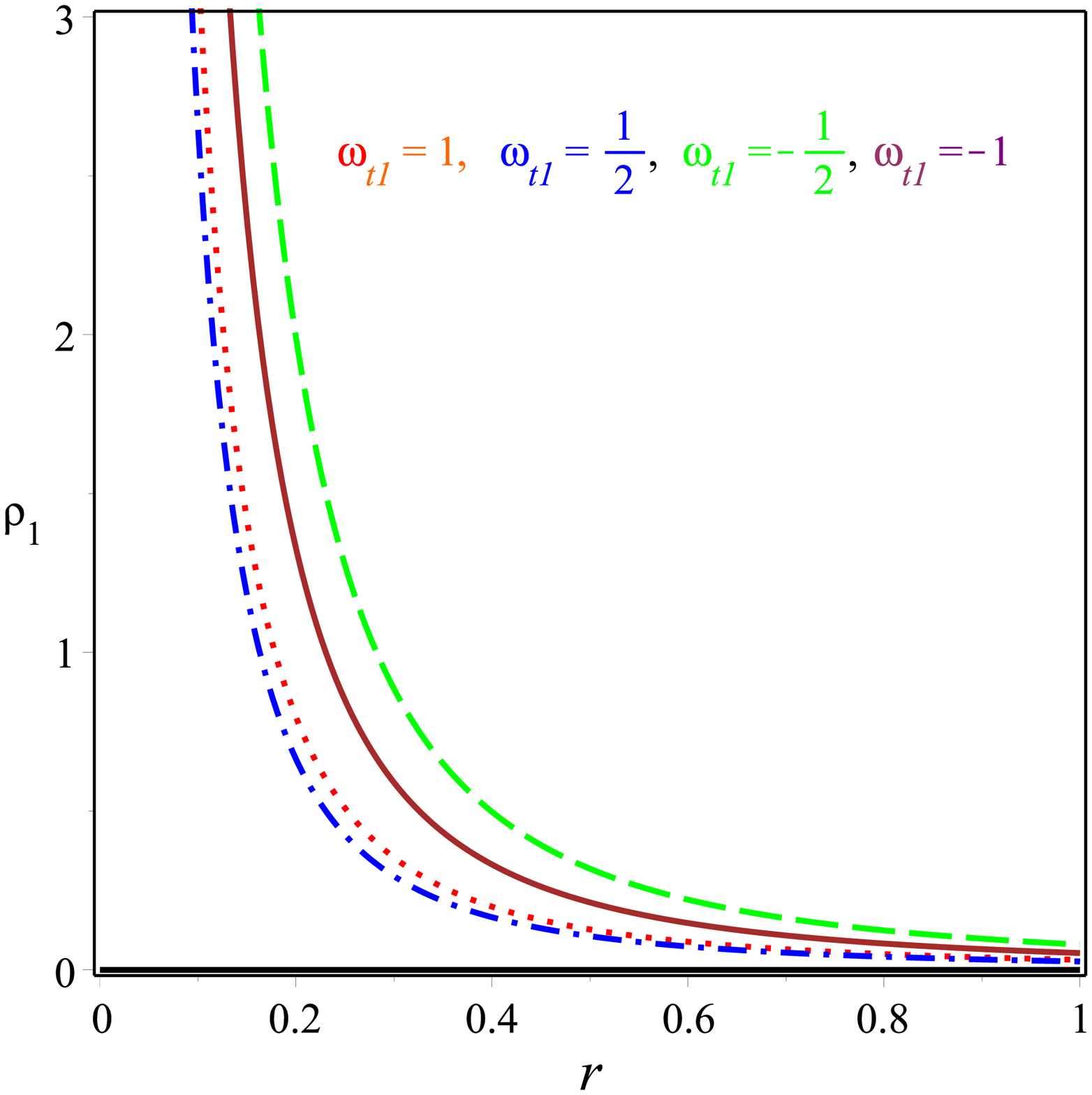}}
\subfigure[~Pressure of the first solution]{\label{fig:pressure}\includegraphics[scale=.3]{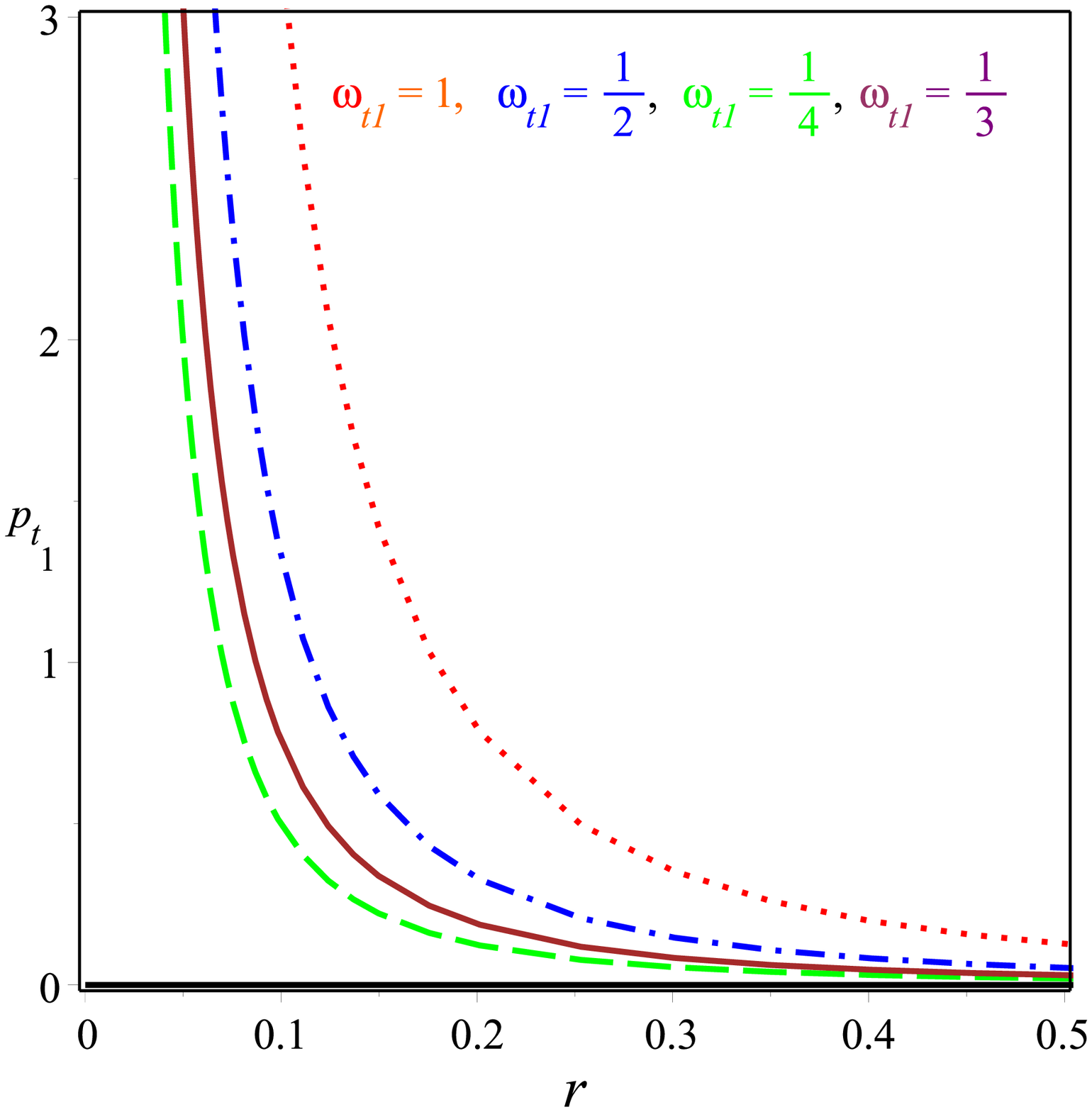}}
\caption[figtopcap]{\small{{Schematic plot of the radial coordinate $r$ in the unit of km versus the energy density and pressure of the solution (\ref{sol}).}}}
\label{Fig:1}
\end{figure}

The behavior of the density and tangential pressure of the first and second solutions are drawn in Fig. \ref{Fig:1}. Figs. \ref{Fig:1}\subref{fig:dnesity} \ref{Fig:1}\subref{fig:pressure} show that energy and pressure decrease as the radial coordinate $r$ increases. For the second solution, Figs.  \ref{Fig:2}\subref{fig:dnesity}, \ref{Fig:2}\subref{fig:pressure} and \ref{Fig:2}\subref{fig:EoS} show that the energy density and pressure become the maximum values for $r=1.5~\mathrm{km} \sim 0.2~\mathrm{km}$ depending on the free parameter $b_0$ and then decreasing\footnote{We vary the value of $b_0$ and leave $b_1$ and $b_2$ fixed because we relate them to mass and radius of the Schwarzschild exterior solution as we will discuss below in the subsection of Matching boundary.}.  As for the EoS parameter of the second solution it takes negative values and then positive values depending also on the value of the free parameter $b_0$. The reason for the variation of the EoS from $negative$ value to $positive$ is because the dominator of Eq. (\ref{es1}) has  only two real solutions that have the form \begin{align}\label{sol2}
  \pm\frac{\sqrt{\sqrt{b_1{}^2-4b_0b_2}-b_1}}{\sqrt{2b_2}}, \qquad \qquad b_2>0.\end{align}
  Equation (\ref{sol2}) ensures that the parameter $b_2$ must not have a zero value and  $b_0<0$ which are consistent with the values given in  Fig.\ref{Fig:2}\subref{fig:EoS} and through the whole of the present study.

 \begin{figure}
\centering
\subfigure[~Density of the second solution]{\label{fig:dnesity}\includegraphics[scale=0.3]{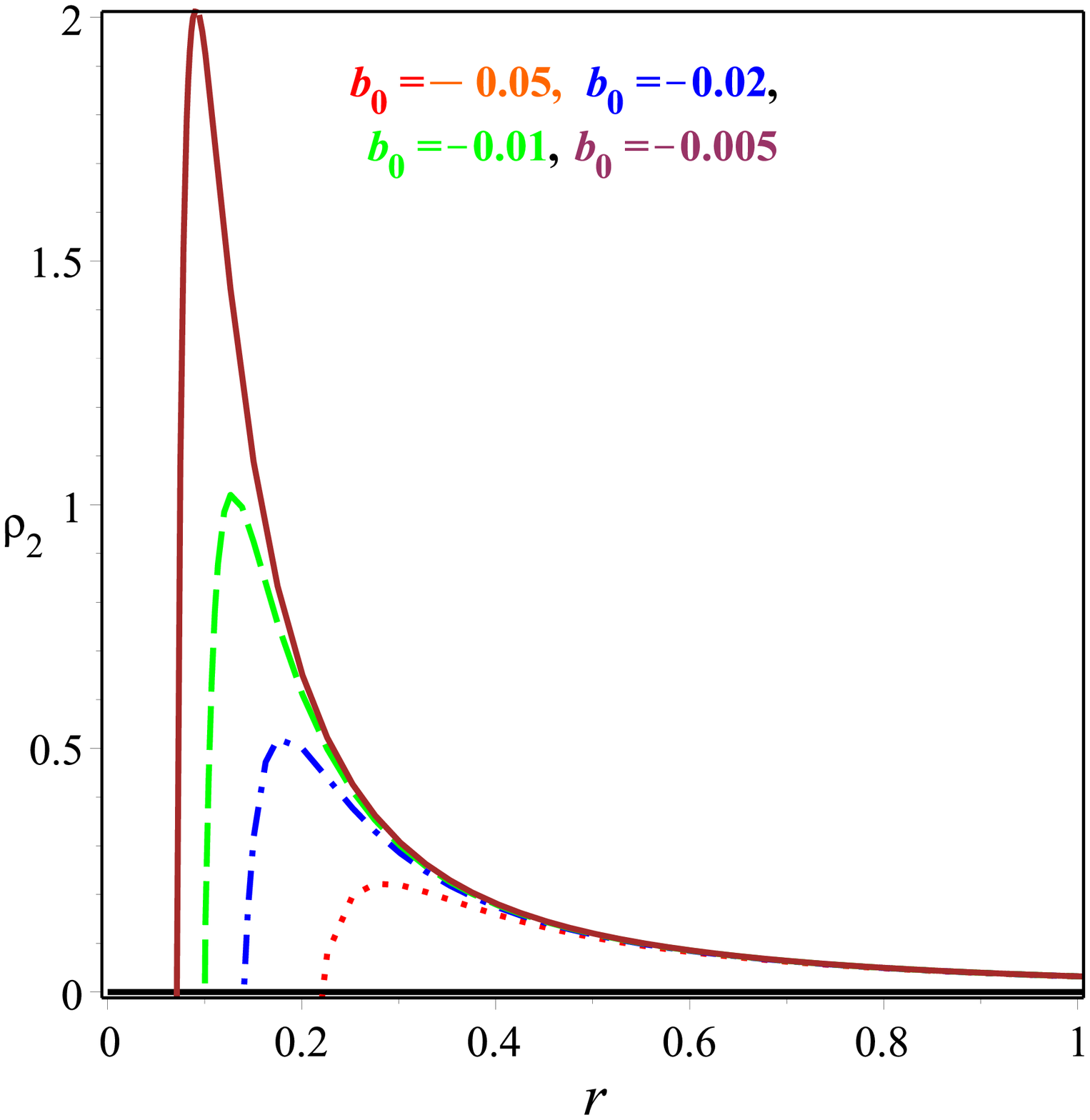}}
\subfigure[~Pressure of the second solution]{\label{fig:pressure}\includegraphics[scale=.3]{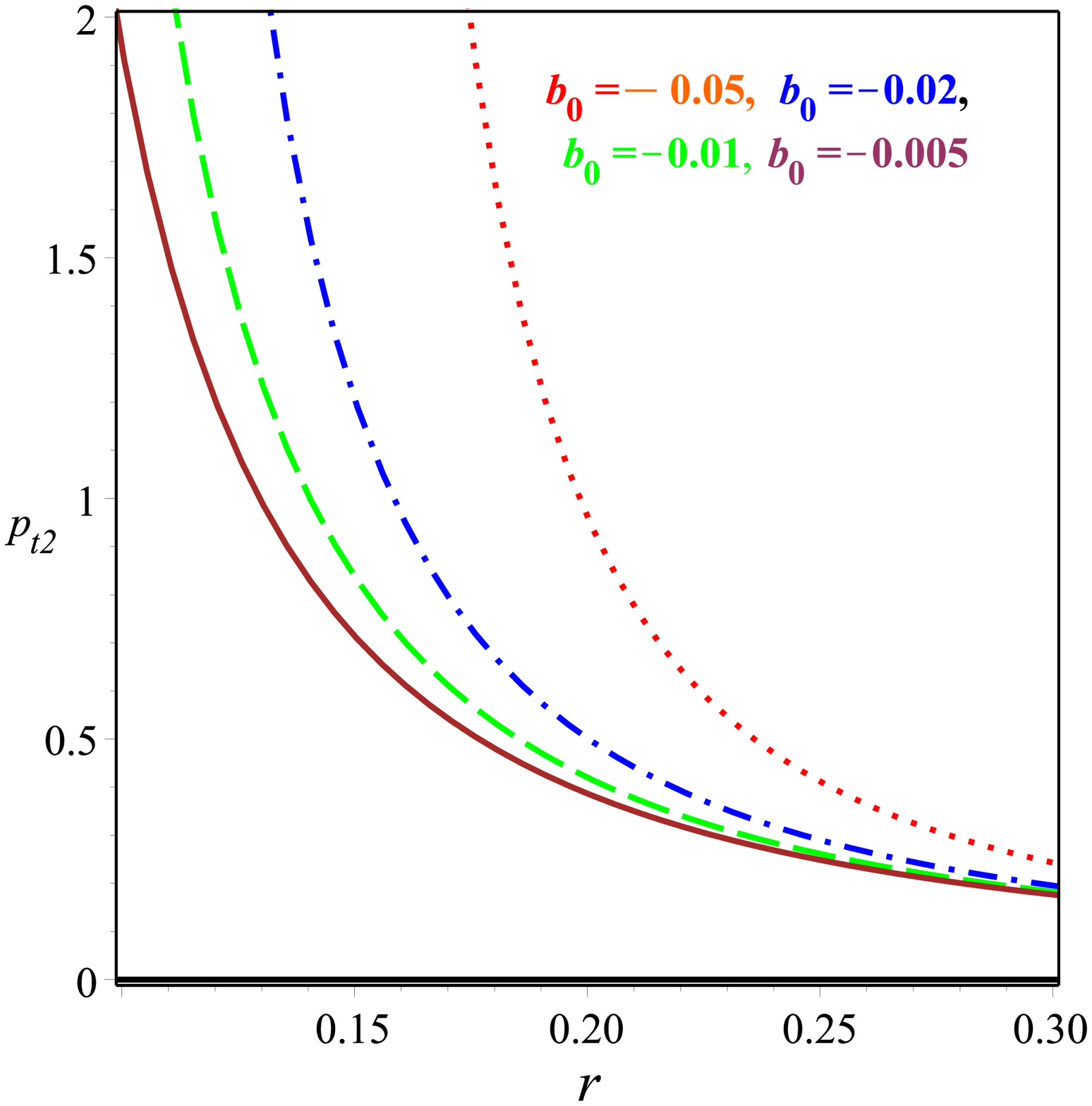}}
\subfigure[~EoS of the second solution]{\label{fig:EoS}\includegraphics[scale=.3]{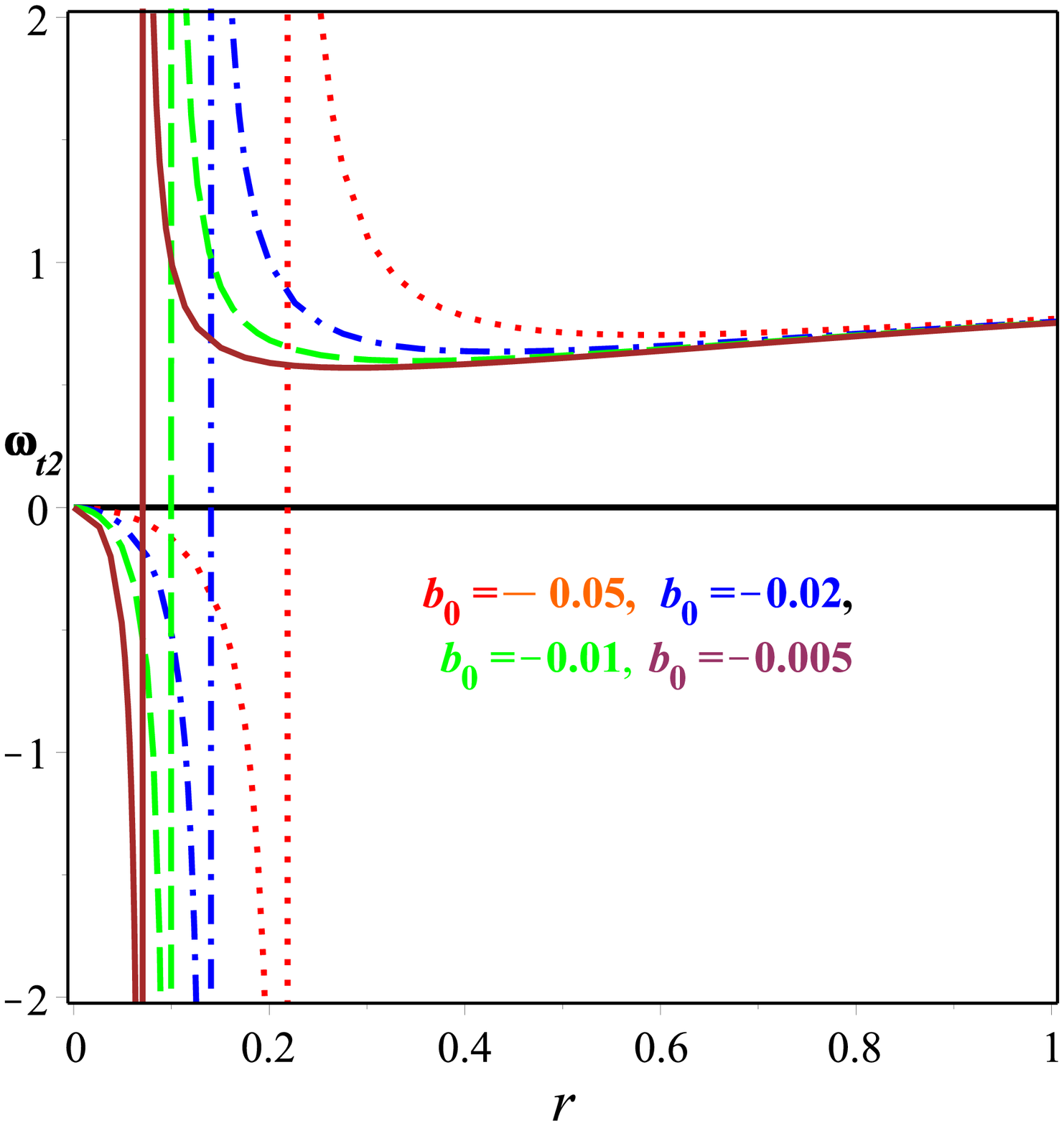}}
\caption[figtopcap]{\small{{Schematic plot of the radial coordinate $r$ in the unit of km versus the energy density, pressure and the EoS of the solution (\ref{sol1}) when  $b_1=b_2=1$.}}}
\label{Fig:2}
\end{figure}

\begin{figure}
\centering
\subfigure[~Mass of solution (\ref{sol}) against radial coordinate]{\label{fig:mass1}\includegraphics[scale=0.3]{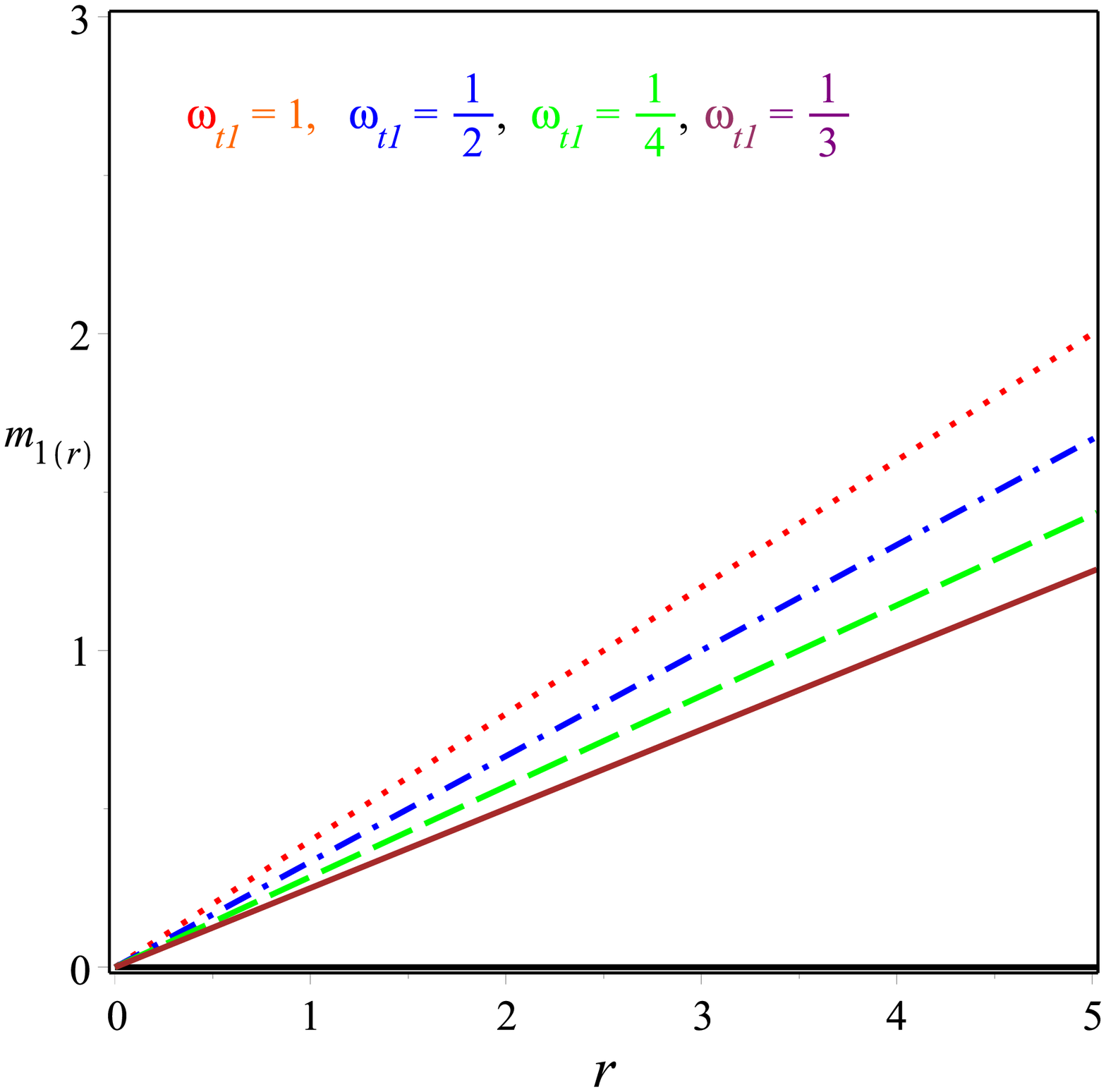}}
\subfigure[~Mass of solution (\ref{sol1}) against radial coordinate]{\label{fig:mass2}\includegraphics[scale=.3]{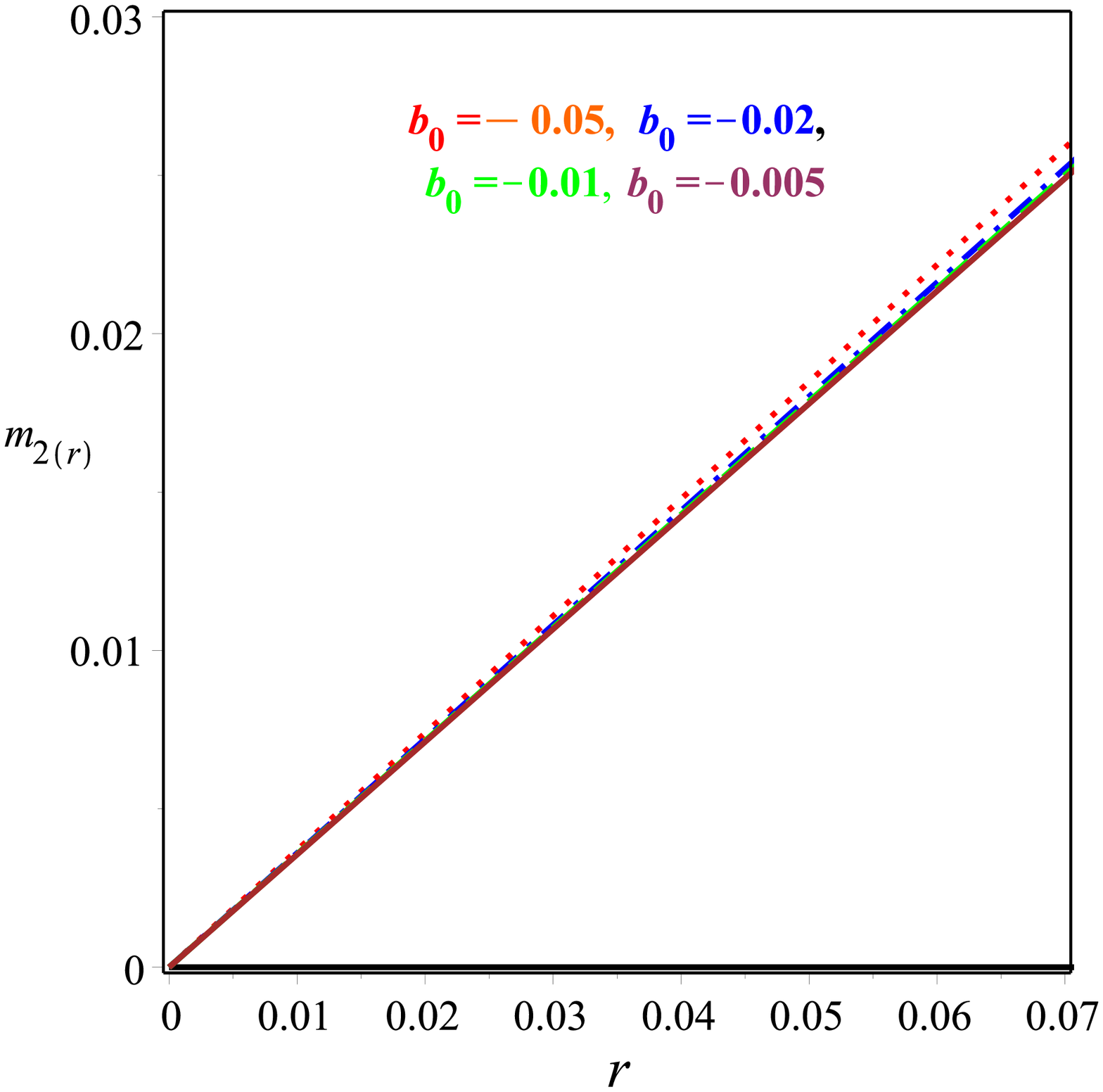}}
\subfigure[~Compactness parameter of solution (\ref{sol1}) against radial coordinate]{\label{fig:U2}\includegraphics[scale=.3]{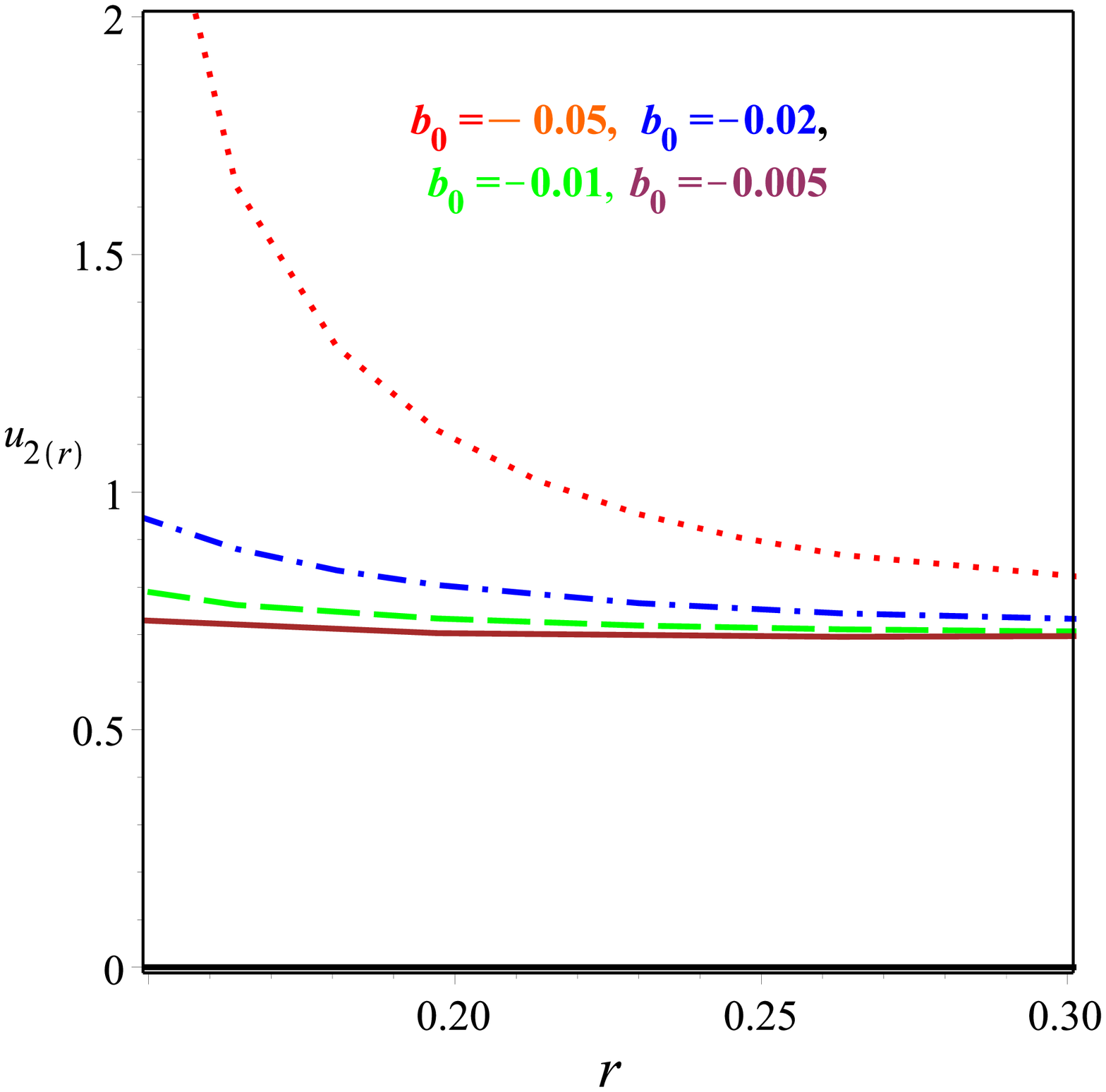}}
\caption[figtopcap]{\small{{Schematic plot of the gravitational mass of solutions (\ref{sol}) and (\ref{sol1}) and  compactness parameter versus the radial coordinate  $r$ in km when  $b_1=b_2=1$.}}}
\label{Fig:3}
\end{figure}

We consider the physical contents for the first and second solutions. To this end, we are going to calculate the following quantities.
The surface red-shift of the first and second solutions  takes the form:
\begin{align}\label{sr}
z_{s_{1}}=e^{\nu_s}-1=4\omega_{t_{1}}\;, \qquad \qquad z_{s_{2}}=\frac{2r^2(b_1+2b_2r^2)}{b_0+b_1r^2+b_2r^4}\;.
\end{align}
The behavior of the surface red-shift of the second solution is identical with the behavior of the EoS, as shown in Fig. \ref{Fig:2}\subref{fig:EoS}, because the two forms are identical up to some constant.
The gravitational mass of a spherically symmetric source with the radial dependence $r$ is expressed by \cite{Singh:2019ykp}
\begin{eqnarray}\label{gm1}
&&m(r)=4\pi{\int_0}^r \rho(\xi) \xi^2 d\xi\;,
\end{eqnarray}
which gives for  solutions (\ref{sol}) and  (\ref{sol1}) the form
\begin{eqnarray}\label{gm}
&&m_1(r)=\frac{2\omega_{t_{1}} r}{4\omega_{t{1}}+1}\,, \qquad \qquad m_2(r)=\frac{20b_2r^3(b_1+2b_2 r^2)}{(3b_1+10b_2r^2)^2-9b_1{}^2+20b_0b_2}\;.
\end{eqnarray}
The behavior of the gravitational mass of solutions (\ref{sol}) and (\ref{sol1}) are drawn  in Fig. \ref{Fig:3}\subref{fig:mass1} and \ref{Fig:3}\subref{fig:mass2}. This figures show the gravitational mass increases with the radial coordinate.
The compactness parameter of a source with its spherical symmetry
in terms of the radius $r$ takes the form \cite{Singh:2019ykp}
\begin{eqnarray}\label{gm1}
&&u(r)=\frac{2m(r)}{r}\, , \qquad \qquad \textrm{that\, gives\, for\,  solutions\, (\ref{sol})\,  and\,  (\ref{sol1})\, the\, form} \nonumber\\
&&u_1(r)=\frac{4\omega_{t_{1}}}{4\omega_{t_{1}}+1}\, , \qquad \qquad u_2(r)=\frac{20r^2(b_1+2b_2 r^2)}{(3b_0+10b_2r^2)^2-9a1^2+20b_0b_2}.
\end{eqnarray}
We show the behavior of compactness parameter of solution (\ref{sol1}), because solution (\ref{sol}) gives a constant value, in Fig. \ref{Fig:3}\subref{fig:U2} which shows some kind of inverse relation, i.e., when $r$ increase $u$ decreases.

The gradient of density and pressure of (\ref{sol}) and (\ref{sol1}) take the form \cite{Singh:2019ykp}
\begin{eqnarray}\label{gm}
&&d\rho_1=-\frac{\omega_t}{2\pi r^3(1+4\omega_t)}\, ,\qquad \qquad \qquad dp_{{t_1}}=-\frac{\omega^2_t}{2\pi r^3(1+4\omega_t)}\, ,\nonumber\\
&&d\rho_2=-\frac{r(10b_0^2b_2-8b_0b_2r^2[8b_1+15b_2r^2]-5b_1b_2r^4[9b_1+20b_2r^2]-15b_0b_2^2-r^2[9b_1^3+50b_2^3r^6])}{2\pi (b_0+3b_1r^2+5b_2r^4)^3}\, ,\nonumber\\
&&dp_{{t_2}}=\frac{r(3b_0[b_1^2+20b_2^2r^4]-9b_1^2r^2[b_1+5b_2r^2]-100b_2{}^2r^6[b_1+b_2r^2]+32b_0b_1b_2r^2)}{4\pi (b_0+3b_1r^2+5b_2r^4)^3}\, .\nonumber\\
\end{eqnarray}
 \begin{figure}
\centering
\subfigure[~Variation of density (\ref{sol}) against radial coordinate]{\label{fig:drho1}\includegraphics[scale=0.22]{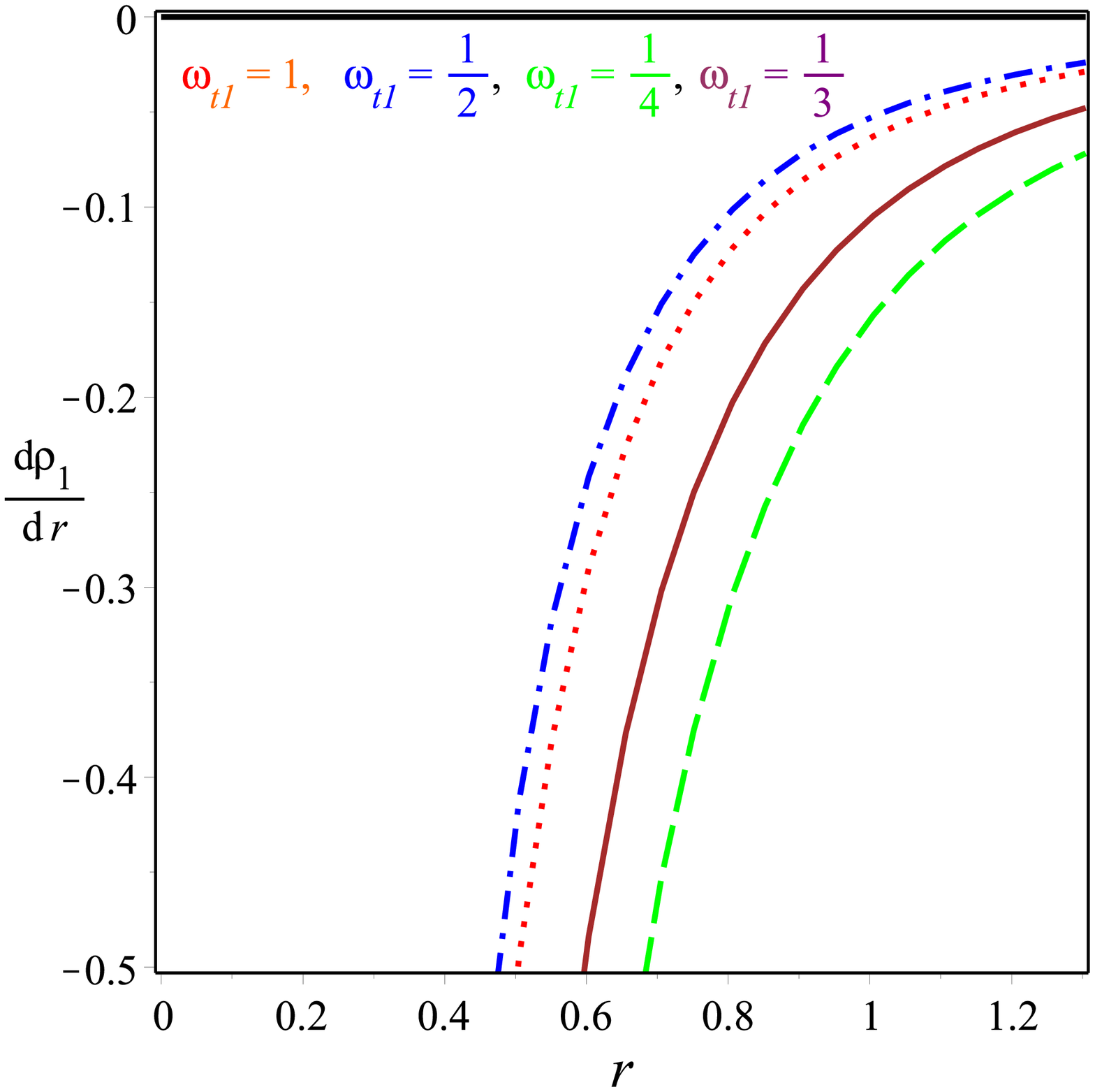}}
\subfigure[~Variation of pressure (\ref{sol}) against radial coordinate]{\label{fig:dp1}\includegraphics[scale=.22]{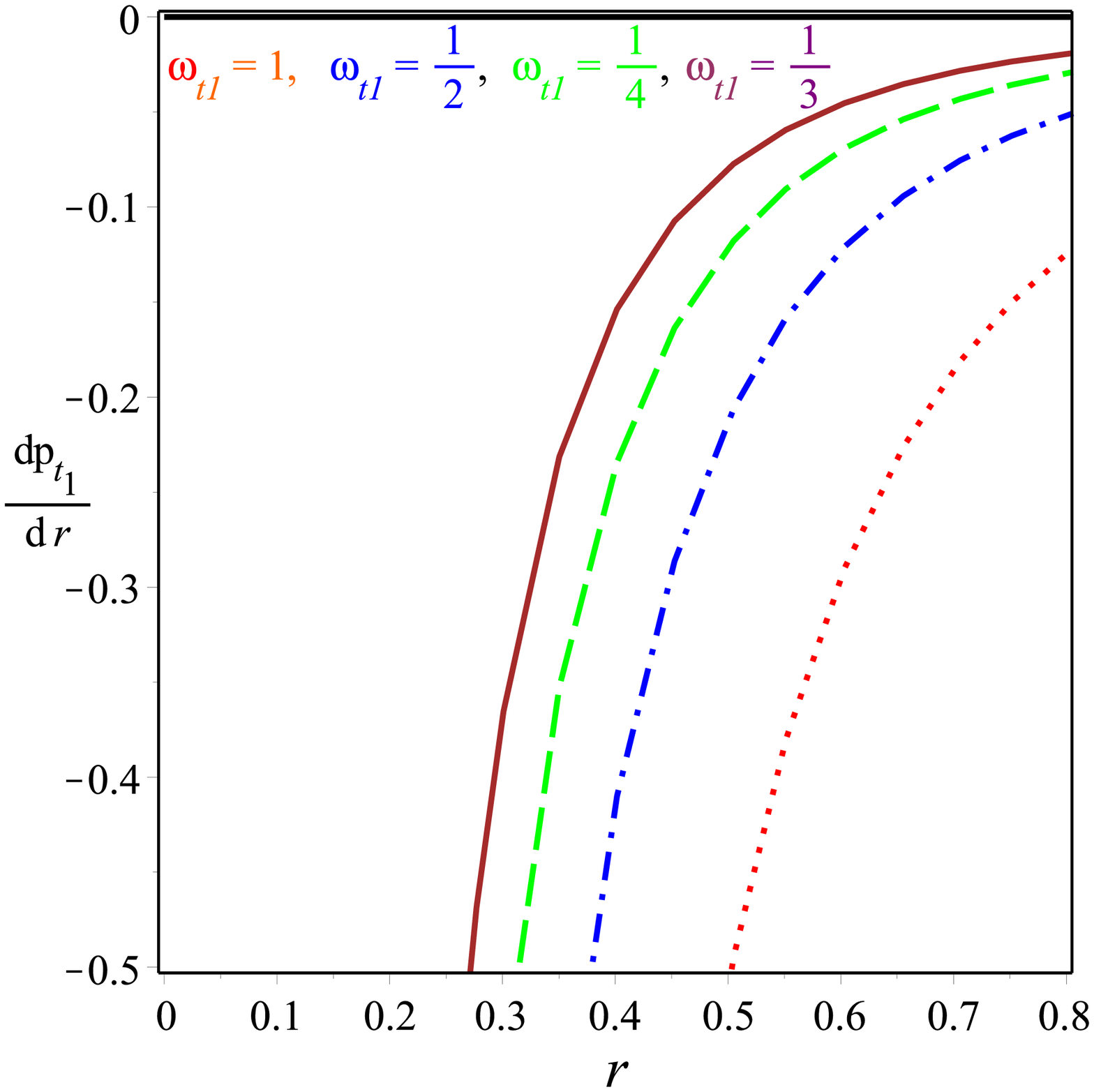}}
\subfigure[~Variation of density (\ref{sol1}) against radial coordinate]{\label{fig:drho2}\includegraphics[scale=0.22]{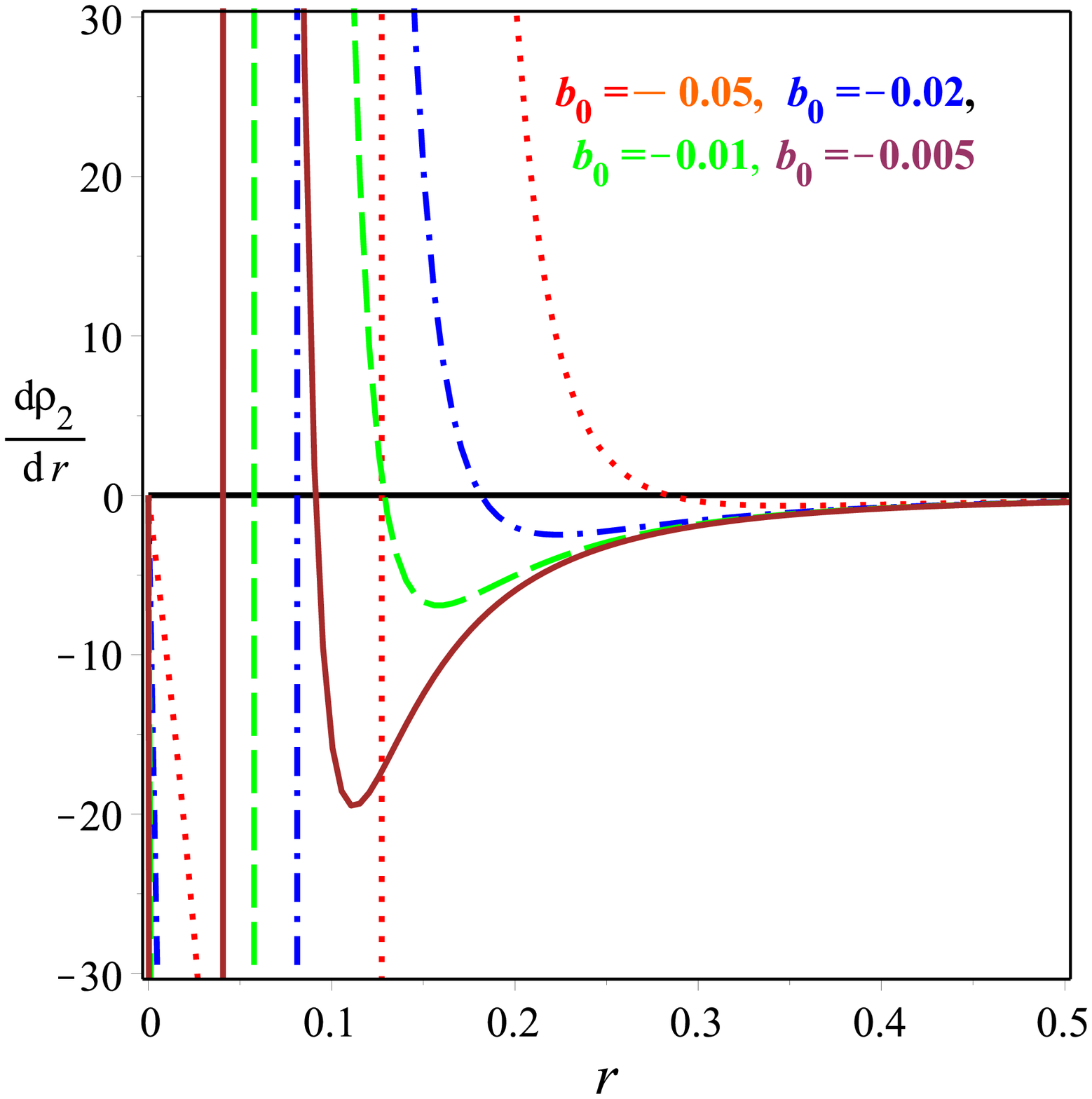}}
\subfigure[~Variation of pressure (\ref{sol1}) against radial coordinate]{\label{fig:dp2}\includegraphics[scale=.22]{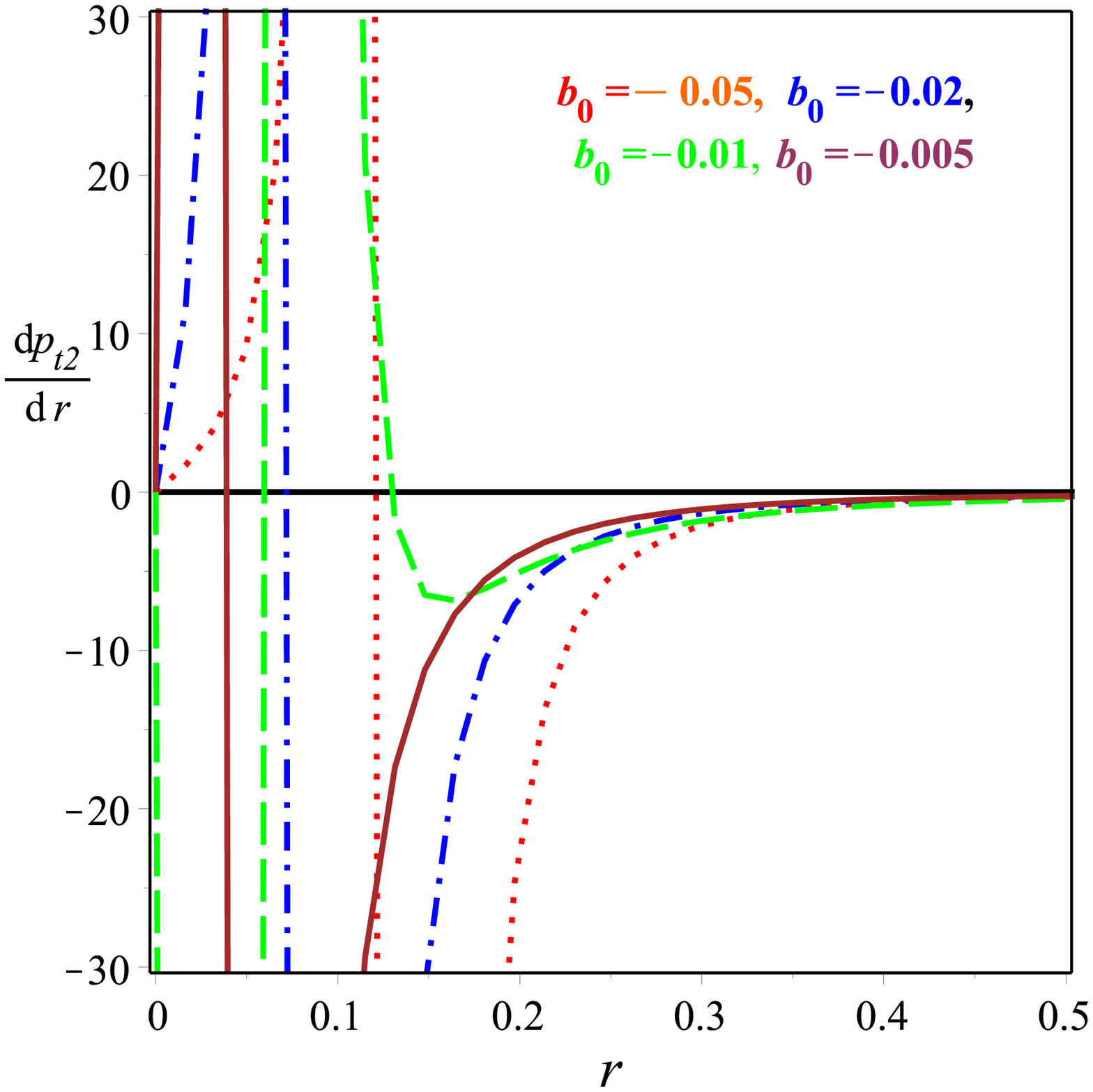}}
\caption[figtopcap]{\small{{Variation of the gradient of density and pressure of (\ref{sol}) and (\ref{sol1}) against  $r$ in km when  $b_1=b_2=1$.}}}
\label{Fig:4}
\end{figure}
Figure \ref{Fig:4} shows that for solution (\ref{sol}) we have always negative gradient for density and pressure while for solution (\ref{sol1}) we have negative value of the gradient of density then this negative changes to positive value and then become negative forever. The change of the sign of density occurs because the dominator of Eq. (\ref{gm}), i.e., $(b_0+3b_1r^2+5b_2r^4)$ has two real solutions $\pm\frac{\sqrt{b_2(\sqrt{9b_1{}^2-20b_0b_2}-3b_1)}}{\sqrt{10b_2}}$ which again ensure that the parameter $b_2\neq 0$ and $b_0<0$. Same discussion can be applied to the gradient of pressure.
\newpage
Finally, the speed of sound of (\ref{sol}) and (\ref{sol1}) take the form \cite{Singh:2019ykp}
\begin{eqnarray}\label{gm2}
&&v_t^2=\frac{dp_{{t}}}{d\rho}\, ,\qquad \textrm{that\, gives\,  for\,  solutions\, (\ref{sol})\,  and\,  (\ref{sol1})\, the\, form} \nonumber\\
&& v_{{t_1}}^2=\omega_t\, , \qquad v_{{t_2}}^2=\frac{3b_0[b_1^2+20b_2^2r^4]-9b_1^2r^2[b_1+5b_2r^2]-100b_2{}^2r^6[b_1+b_2r^2]+32b_0b_1b_2r^2}
{10b_0^2b_2-8b_0b_2r^2[8b_1+15b_2r^2]-5b_1b_2r^4[9b_1+20b_2r^2]-15b_0b_2^2-r^2[9b_1^3+50b_2^3r^6]}\, .\nonumber\\
\end{eqnarray}
 \begin{figure}
\centering
{\label{fig:v2}\includegraphics[scale=0.4]{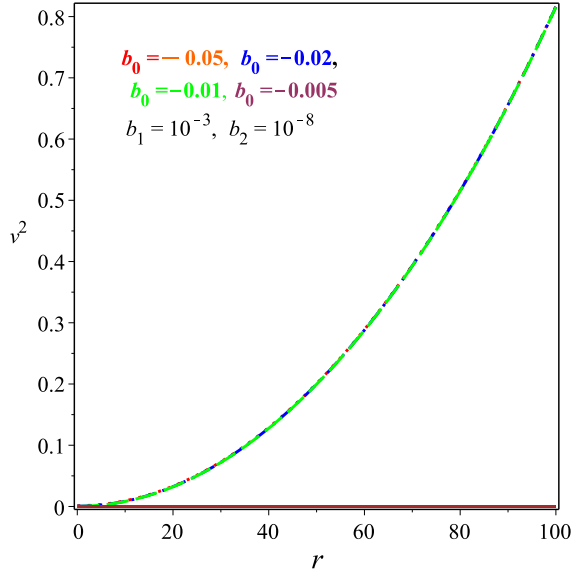}}
\caption[figtopcap]{\small{{Speed of sound  of (\ref{sol1})  against  $r$ in km.}}}
\label{Fig:5}
\end{figure}
We discuss the property of the speed of sound in the second solution because the first one gives a constant, which depends on the EoS parameter. Usually, the sound velocity must be less than the light speed \cite{Singh:2019ykp}. Hence, in relativistic units, the sound speed must be less than or equal to unity. Thus, for the first solution, to give the sound speed less than or equal to unity, we must have $\omega_t \leq1$. As Fig. \ref{Fig:5} shows  for solution (\ref{sol1}), we have speed of sound less than 1 when the parameters $b_1=10^{-4}$ and $b_2=10^{-8}$.

\section{Physics of the  compact stars (\ref{sol}) and (\ref{sol1}) }\label{S4}
In this section, we explore the physical consequences for the first and second solutions given by Eqs. (\ref{sol}) and (\ref{sol1}). To this end, first we are going to determine the values of the constants appearing in these solutions.

\subsection{Matching of boundary }
We compare the solution within the compact objects with the Schwarzschild vacuum solution outside it. We use the first solution in Eq. (\ref{sol}) with the Schwarzschild one, i.e.,
\begin{eqnarray}\label{Eq1}  ds^2= -\Big(1-\frac{2M}{r}\Big)dt^2+\Big(1-\frac{2M}{r}\Big)^{-1}dr^2+r^2d\Omega^2.
\end{eqnarray}
This yields the following matching conditions:
\begin{eqnarray} && 1-\frac{2M}{\cal{R}}=4\omt\ln({\cal{R}})+c_1\, ,\nonumber\\
&& \frac{1}{1-\frac{2M}{\cal{R}}}=\ln(4\omt+1)\, ,\label{Eq1}
 \end{eqnarray}
 where ${\cal{R}}$ is the radius at the boundary, i.e., at the boundary  $r ={\cal{R}}$.
Solving for $\omt$ and $c_1$ from Eq. (\ref{Eq1}), we obtain
\begin{eqnarray} && \omega_t = \frac{e^{\frac{1}{1-\frac{2M}{\cal{R}}}}-1}{4}\, ,\qquad   \textrm {and } \qquad  c_1=\frac{{\cal{R}}-2M+{\cal{R}}\ln\,{\cal{R}} -{\cal{R}}\ln\, {\cal{R}} {e^{\frac{1}{(1-\frac{2M}{\cal{R}})}}}}{\cal{R}}.
\label{com1}
\end{eqnarray}
Here $M$ and ${\cal{R}}$ are determined by the observations of the compact objects. Applying the same procedure to the second solution (\ref{sol1}) we get
 \begin{eqnarray} && b_1 = \frac{2{\cal{R}}(1-b_0)+5M}{{\cal{R}}^3}\, ,\qquad  \qquad   b_2= \frac{3M-{\cal{R}}(1+b_0)+5M}{{\cal{R}}^5}\,,\label{com1}
 \end{eqnarray}
where $b_0$ is tackled by the data fitting and the values of $M$ and ${\cal{R}}$ are selected from the observations of the compact objects.

\subsection{Energy conditions for compact stars}
\begin{figure}
\centering
\subfigure[~Energy conditions of solution (\ref{sol}) against radial coordinate]{\label{fig:en}\includegraphics[scale=0.35]{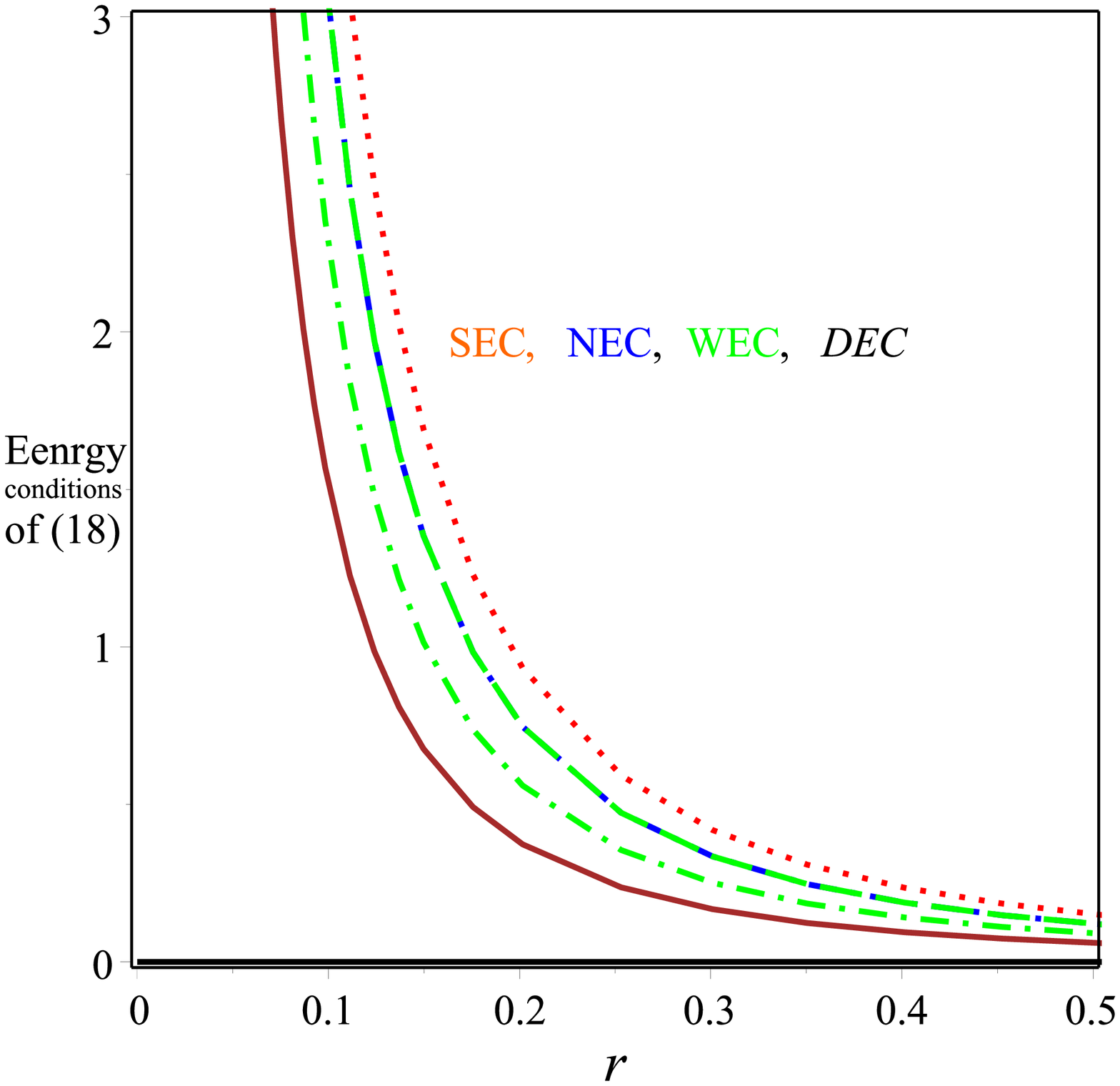}}
\subfigure[~Energy conditions of solution (\ref{sol1}) against radial coordinate]{\label{fig:dp1}\includegraphics[scale=.35]{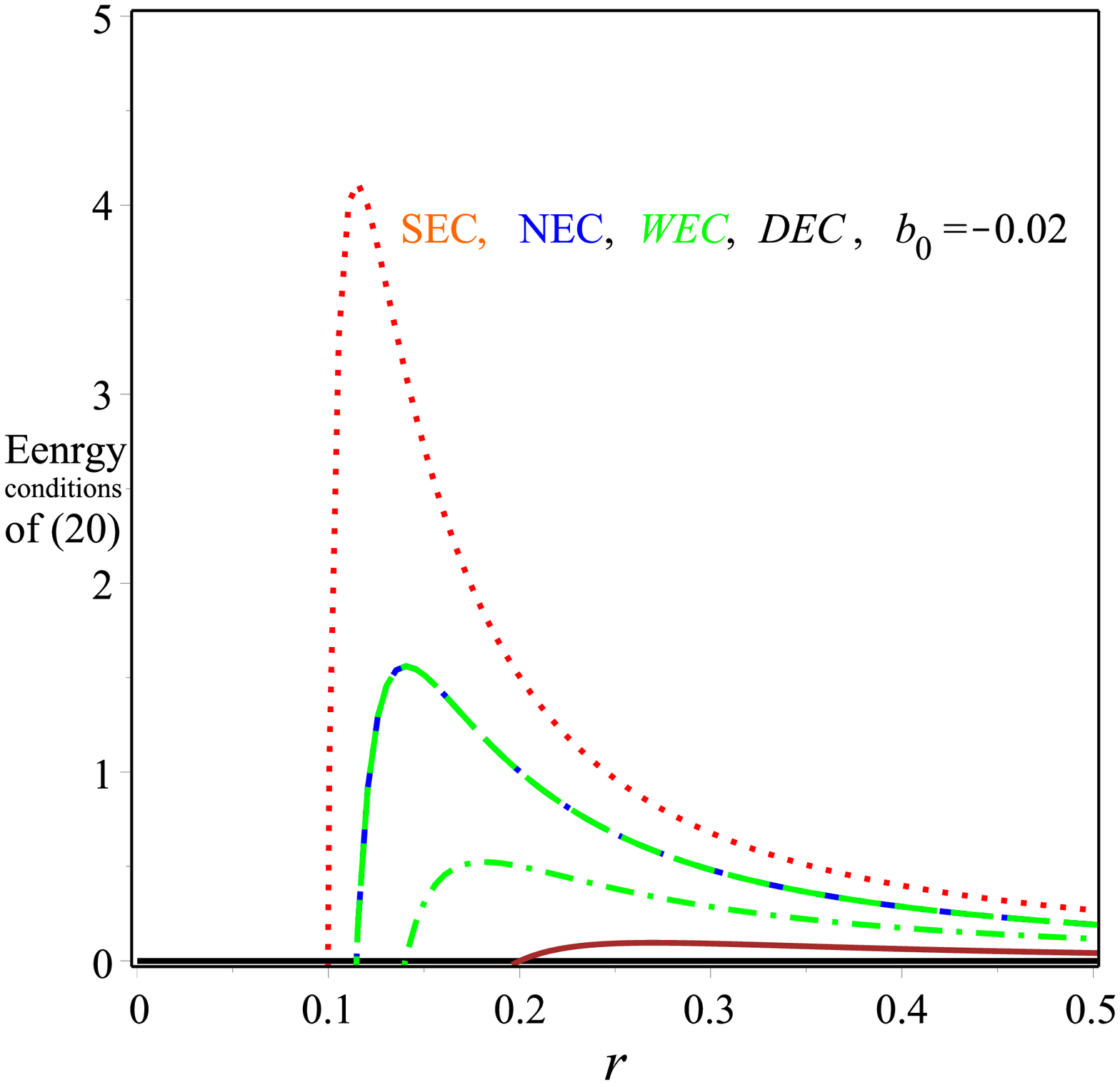}}
\caption[figtopcap]{\small{{The energy conditions of  solutions (\ref{sol}) and (\ref{sol1}) against  $r$ in km when  $b_1=b_2=1$.}}}
\label{Fig:6}
\end{figure}

In general, for perfect fluid models, the energy conditions described by
the relation between the energy density and pressure can be satisfied.
We check strong (SEC), weak (WEC), dominant (DEC) and finally null (NEC) energy conditions, given by
\begin{eqnarray}\label{ec}
&&\textrm{SEC}:  \rho+2p_t\geq 0\,, \qquad \qquad \qquad \qquad \textrm{NEC}:  \rho+p_t\geq 0\,,\nonumber\\
&&\textrm{WEC}:  \rho\geq0, \qquad \rho+p_t\geq 0\,, \qquad \qquad \textrm{DEC}:  \rho\geq \mid p_t\mid\,.\nonumber\\
\end{eqnarray}
By using Eqs. (\ref{sol}) and (\ref{sol1}), one can easily show that the above conditions are satisfied as indicated in Figs. \ref{Fig:6} \subref{fig:en} and \subref{fig:dp1}.

\subsection{Tolman-Oppenheimer-Volkoff equation and the analyses of the equilibrium}
In this subsection we are going to discuss

\begin{figure}
\centering
\subfigure[~TOV of solution (\ref{sol}) against radial coordinate]{\label{fig:TOV1}\includegraphics[scale=0.35]{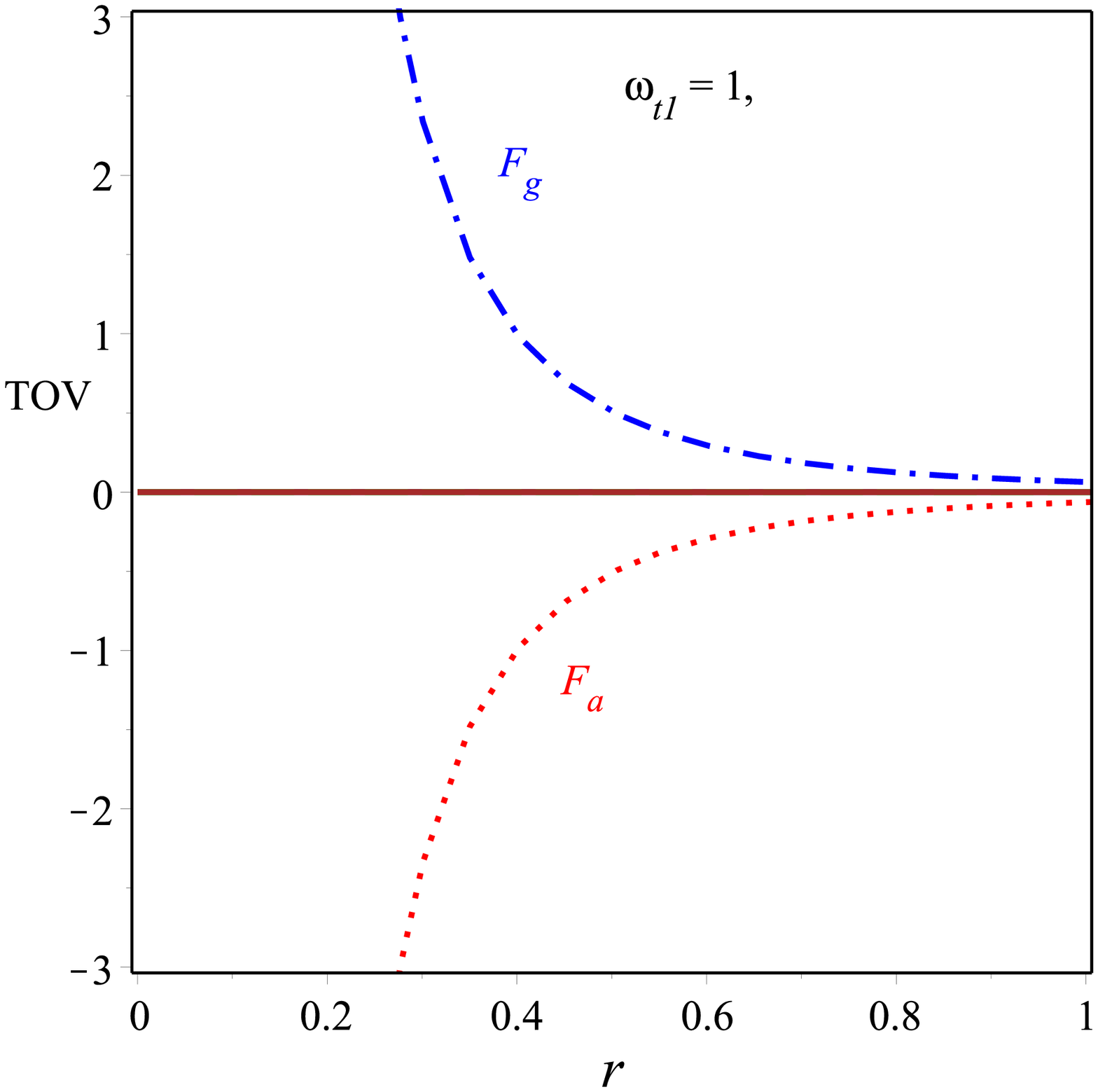}}
\subfigure[~TOV of solution (\ref{sol1}) against radial coordinate]{\label{fig:TOV2}\includegraphics[scale=.35]{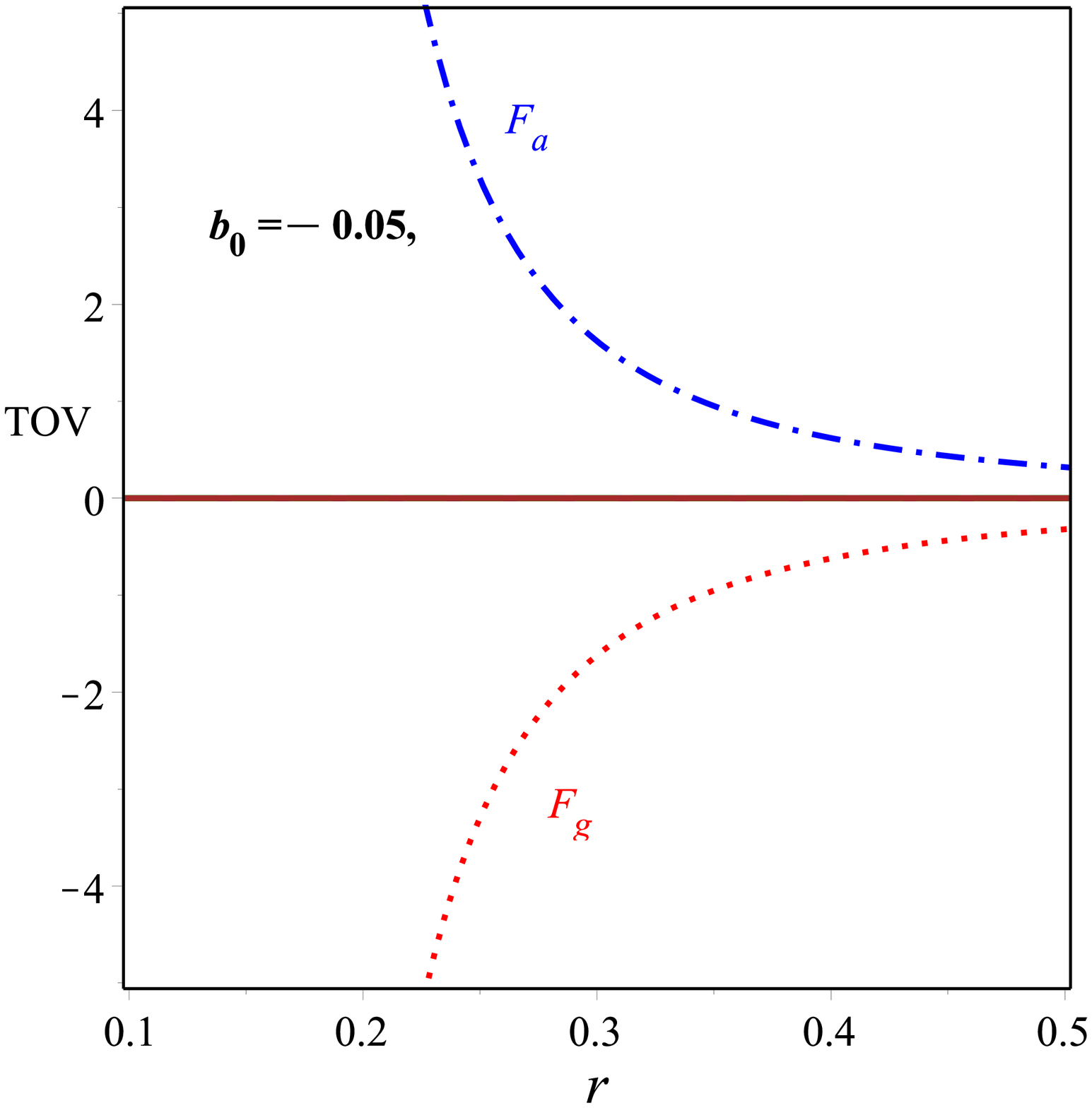}}
\caption[figtopcap]{\small{{TOV solutions (\ref{sol}) and (\ref{sol1}) against  $r$ in km when  $b_1=b_2=1$.}}}
\label{Fig:7}
\end{figure}

We investigate how stable the models of the Einstein's clusters are.
We assume the equilibrium of the hydrostatic state.
Through the Tolman-Oppenheimer-Volkoff (TOV) equation \cite{PhysRev.55.364,PhysRev.55.374} as that presented in \cite{PoncedeLeon1993}, we acquire the equation
\begin{eqnarray}\label{TOV}   \frac{2p_t(r)}{r}-\frac{M_g(r)\rho(r)e^{(\mu-\nu)/2}}{r}=0\;,
 \end{eqnarray}
with $M_g(r)$ the gravity mass as a function of $r$,
which is defined by the Tolman-Whittaker mass formula as
\begin{eqnarray}\label{ma}   M_g(r)=4\pi{\int_0}^r\Big({T_t}^t-{T_r}^r-{T_\theta}^\theta-{T_\phi}^\phi\Big)r^2e^{(\mu+\nu)/2}dr=\frac{re^{(\nu-\mu)/2}\mu'}{2}\,,
 \end{eqnarray}
Using Eq. (\ref{ma}) in (\ref{TOV}) we get
\begin{eqnarray}\label{ma1}  \frac{2p_t(r)}{r}-\frac{\mu'\rho(r)}{2}=F_g+F_a=0\,,
 \end{eqnarray}
 with $F_g=-\frac{\mu'\rho(r)}{2}$ being the gravitational force and $F_a=\frac{2p_t(r)}{r}$ is the anisotropic force. The behaviors of the TOV equations of solutions (\ref{sol}) and (\ref{sol1}) are shown in Fig. \ref{Fig:7} \subref{fig:TOV1} and \subref{fig:TOV2}, respectively.

\subsection{Relativistic adiabatic index and stability analysis}
\begin{figure}
\centering
{\label{fig:ad}\includegraphics[scale=.4]{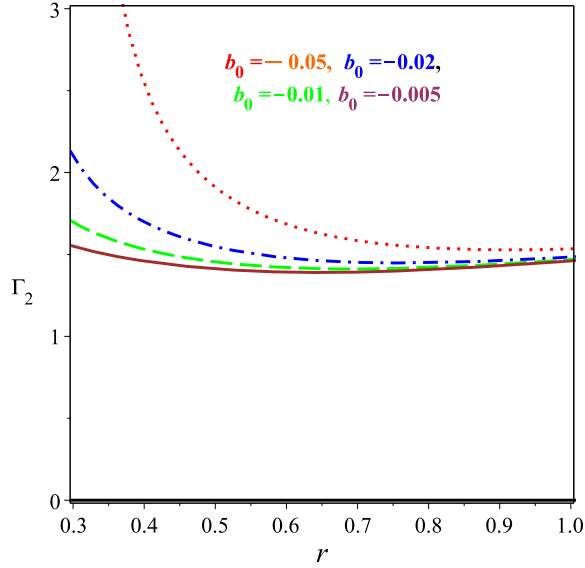}}
\caption[figtopcap]{\small{{Adiabatic index of  (\ref{sol1}) against  $r$ in km when  $b_1=b_2=1$.}}}
\label{Fig:8}
\end{figure}

Our particular interest is to study the stable equilibrium
configuration of a spherically symmetric cluster, and the adiabatic index is a basic ingredient of
the stable/unstable criterion. Now considering an adiabatic
perturbation, the adiabatic index $\Gamma$ is defined as \cite{1964ApJ...140..417C,1989A&A...221....4M,10.1093/mnras/265.3.533}
\begin{eqnarray}\label{ai}  \Gamma=\frac{\rho+p_t}{p_t}\frac{dp_t}{d\rho}\,,
 \end{eqnarray}
 with $\frac{dp_t}{d\rho}$ is the speed of sound.
 Using Eq. (\ref{ai}) we get the adiabatic index of the two solutions (\ref{sol}) and (\ref{sol1}) in the form:
 \begin{eqnarray}\label{aic} && \Gamma_1=1+\omega_{t1}\,, \nonumber\\
 &&\Gamma_2=\frac{(2b_0+3b_1r^2+4b_2r^4)(3b_0b_1{}^2-45b_1{}^2b_2r^4-100b_1b_2{}^2r^6+32b_0b_1b_2r^2-100b_2{}^3r^8+ 60b_0b_2{}^2r^4-9b_1{}^3r^2)}{2r^2(b_1+2b_2r^2)(10b_0{}^2b_2-64b_0b_1b_2r^2-120b_0b_2{}^2r^4-45b_1{}^2b_2r^4-100b_1b_2{}^2r^6 -50b_2{}^3r^8-15b_0b_1{}^2-9b_1{}^3r^2)}\,. \nonumber\\
 &&
 \end{eqnarray}
The first set of Eq. (\ref{aic}) is always larger than or equal to unity, depending on the value of EoS of $\omega_{t1}$. The behavior of the second set of Eq. (\ref{aic}) is shown in Fig. \ref{Fig:8}. From this figure, we can see that the adiabatic index is always larger than unity and its value depends of the parameters $b_0$, $b_1$ and $b_2$.
It has been found by Bondi \cite{Bondi:1964zza} that in the case of non-charged equilibrium, $\Gamma= 4/3$ for the stable Newtonian sphere. It is shown in \cite{Haensel:2007yy} that the variable range in terms of the value of $\Gamma$ is larger than or equal to 2 and less than or equal to 4 for the equations of state of most of the neutron stars.

\subsection{Stability in the static state}
For stable compact stars, in terms of the mass-central as well as mass-radius relations for the energy density,
Harrison, Zeldovich and Novikov [80, 81] claimed
the gradient of the central density with respect to mass increase must be positive, i.e., $\frac{\partial M}{\partial \rho_{r_0}}> 0$. If this condition is satisfied then we have stable configurations. To be more specific,  stable or unstable region is satisfied
for constant mass i.e. $\frac{\partial M}{\partial \rho_{r_0}}= 0$. Let us apply this procedure to our solutions (\ref{sol}) and (\ref{sol1}). To this end we calculate the  central density for both solutions. For solution (\ref{sol}) the  central density is undefined so we will exclude this case from our consideration because it may represent unstable configuration. As for the second solution  the  central density has the form
\begin{eqnarray} \label{sta} &&\rho_{r_0}=\frac{3b_1}{4\pi b_0} \Rightarrow b_0=\frac{3b_1}{4\pi \rho_{r_0}},\nonumber\\
 && M(\rho_{r_0})=\frac{4\pi R^3\rho_{r_0}  (b_1+2b_2 R^2)}{3b_1+12\pi b_1R^2\rho_{r_0}+20\pi b_2 R^4 \rho_{r_0}} \,.
 \end{eqnarray}
With Eq. (\ref{sta}) we have
 \begin{eqnarray} \label{sta1} \frac{\partial M}{\partial \rho_{r_0}}=\frac{12\pi R^3 b_1(b_1+2b_2 R^2)}{(3b_1+12\pi b_1R^2\rho_{r_0}+20\pi b_2 R^4 \rho_{r_0})^2} \,.
 \end{eqnarray}
 From Eq. (\ref{sta1}), it is seen that the solution (\ref{sol1}) has a stable configuration since  $\frac{\partial M}{\partial \rho_{r_0}}> 0$. The behavior of the adiabatic index is shown in

\begin{figure}
\centering
{\label{fig:st}\includegraphics[scale=.4]{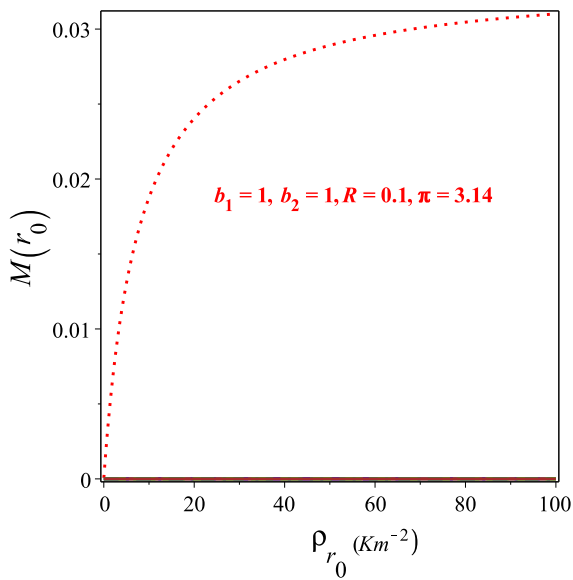}}
\caption[figtopcap]{\small{{Static stability of  (\ref{sol1}) against  $\rho_{r_0}$ in $\mathrm{km}^{-3}$ when $b_1=b_2=1, R=0.1$.}}}
\label{Fig:9}
\end{figure}

Figure \ref{Fig:9} depicts the mass in terms of the energy density.
It follows from this figure that the mass increases as the energy density becomes larger.

\section{Discussions and Conclusions}\label{S5}
In this study, we have explored and discussed the model for compact stars
which mimic clusters for TEGR. The gravitational field equations of the non-vacuum TEGR theory have been applied to a tetrad field with its non-diagonal components, which consists of functions of $\mu$ and $\nu$ possessing spherically symmetric fields. We have derived a set of the three equations with differentiations in terms of the five unknown quantities: $\mu$, $\nu$, $\rho$, $p_r$ and $p_t$. To be able to solve this system, we have put the radial pressure equal to zero \cite{Boehmer:2007az,Singh:2019ykp} in addition to two different assumptions:

\begin{itemize}
\item In our first assumption we have taken an EoS between the density and the tangential pressure in the form $p_t=\omega_{{t_1}} \rho$. By using the vanishing of the pressure in the radial direction and the EoS parameter, we have solved the set of the differential equations and obtained two different solutions. One of these solutions is just the Schwarzschild exterior solution and we excluded it and the other one gave the unknown functions $\mu$, $\nu$, $\rho$ depending on the radial coordinate $r$,  the parameter of EoS $\omega_{{t_1}}$ and on a constant of integration. We have studied the physics of this solution and shown that it has a positive density and pressure and a positive gravitational mass as shown in Figs. \ref{Fig:1} \subref{fig:dnesity}, \ref{Fig:1} \subref{fig:pressure} and \ref{Fig:3} \subref{fig:mass1}. We have found that the speed of sound depends on the the parameter of EoS, $\omega_t$ which should be less than one, i.e., $\omega_t\leq 1$  \cite{Singh:2019ykp}. We have also studied the boundary condition, i.e., matching our solution on the boundary with the exterior Schwarzschild solution, we derived the relations between the EoS parameter, the constant of integration and the gravitational mass of Schwarzschild and its radius at the boundary. Moreover, we showed that this solution satisfies all the energy conditions, i.e.,  SEC, WEC, DEC and NEC.  As shown in  Fig. \ref{Fig:7} \subref{fig:TOV1}, this solution satisfies the TOV equation. Finally, we have demonstrated that the adiabatic index of this solution is satisfied provided that $\omega_t\geq 1/3$ to have $\Gamma \geq 4/3$ \cite{Singh:2019ykp}.

\item In the second assumption, we have used a specific form of the unknown function $\mu$ that has three constants and derived the other unknown functions $\nu$, $\rho$ ad $p_t$. We have repeated the above procedure and shown that this solution has a positive density, a positive tangential pressure and a positive gravitational mass as shown in Figs. \ref{Fig:2} \subref{fig:dnesity}, \ref{Fig:2} \subref{fig:pressure} and \ref{Fig:3} \subref{fig:mass2}. Also we have found that the sound speed depends on the radial coordinate and is always less than 1 as indicated in Fig. \ref{Fig:5}. The energy conditions of this solution are satisfied as shown in Fig. \ref{Fig:6} \subref{fig:dp1}. We have matched our solution with the Schwarzschild exterior and derived a relation between two constants that characterize the unknown function $\mu$ with the gravitational mass and boundary radius of Schwarzschild and dealt with the third constant as a  fitting parameter. Moreover, we showed that this solution satisfies the TOV equation as shown in  Fig. \ref{Fig:7} \subref{fig:TOV2}. We have illustrated that the adiabatic index of this solution is satisfied and always has $\Gamma \geq 4/3$ as drawn in Fig. \ref{Fig:8} \cite{Singh:2019ykp}. Finally, we have demonstrated that the static stability is always satisfied because the derivative of the gravitational mass w.r.t. central density is always positive, indicating the gravitational mass increases with the central density as shown in Fig. \ref{Fig:9}.
\end{itemize}

To summarize, in the present paper we have used a non-diagonal form of tetrad field that gives null value of the off diagonal components of  the field equations unlike what has been studied in the literature \cite{PhysRevD.98.064047,Abbas:2015yma,Momeni:2016oai,2015Ap&SS.359...57A,Chanda:2019hyh,Debnath:2019yge,Ilijic:2018ulf}. The results of this study give satisfactory physical compact stars as shown in the above discussion.

\section*{Acknowledgments}
AA acknowledges that this work is based on the research supported in part by the
National Research Foundation (NRF) of South Africa (grant numbers 109257 and 112131). The work of KB has been partially supported by the JSPS KAKENHI Grant Number JP
25800136 and Competitive Research Funds for Fukushima University Faculty (19RI017). AA acknowledges the hospitality of the High Energy and Astroparticle Physics Group of the Department of Physics of Sultan Qaboos University, where part of this work was completed.

%

\end{document}